\documentclass[traditabstract]{aa} % for the abstract without structuration 

\usepackage{graphicx}
\usepackage{subfig}
\usepackage{natbib}
\usepackage{longtable}
\usepackage[english]{babel}
\usepackage{txfonts}
\def \xmm {\emph{XMM-Newton}}
\def \chandra {\emph{Chandra}}
\def \deg {$^{\circ}$}
\def \rosat {\emph{ROSAT}}

\begin{document}
 \title{Abell 2142 at large scales: An extreme case for sloshing?}
\author{Rossetti M.\inst{1,2}
\and Eckert D.\inst{3,2}
\and De Grandi S.\inst{4}
\and Gastaldello F. \inst{2,5}
\and Ghizzardi S. \inst{2}
\and Roediger E. \inst{6}
\and Molendi S. \inst{2}
}
\institute{ Universit\`a degli Studi di Milano, Dip.~di Fisica, via Celoria 16, 20133 Milano, Italy
\and IASF-Milano INAF, via Bassini 15, 20133, Milano, Italy
\and Department of Astronomy, University of Geneva, ch.~d'Ecogia 16, 1290 Versoix, Switzerland
\and INAF, Osservatorio Astronomico di Brera, via E.~Bianchi 46, 23807, Merate (LC), Italy
\and Department of Physics and Astronomy, University of California at Irvine, 4129 Frederick Reines Hall, Irvine, CA 92697-4572, USA
\and Hamburger Sternwarte, Universitaet Hamburg, Gojensbergsweg 112, D-21029 Hamburg, Germany 
}
\abstract{ We present results obtained with a new \xmm\ observation of A2142, a textbook example of a cluster with multiple cold fronts, which has been studied in detail with \chandra\ but whose large scale properties are presented here for the first time. We report the discovery of 
a new cold front, the most distant one ever detected in a galaxy cluster, at about 1 Mpc from the center to the SE. Residual images, thermodynamics, and metal abundance maps qualitatively agree with predictions from numerical simulations of the sloshing phenomenon. However, the scales involved are much larger, similar to what has been recently observed in the Perseus cluster. These results show that  sloshing is a cluster-wide phenomenon and is not confined in the cores. Sloshing extends well beyond the cooling region, involving a high fraction of the ICM up to almost half of the virial radius. The absence of a cool core and a newly discovered giant radio halo in A2142, in spite of its relaxed X-ray morphology, suggest that large scale sloshing, or the intermediate merger that caused it, may trigger Mpc-scale radio emission and may lead to the disruption of the cluster cool core.}
\keywords{Galaxies:clusters:general--Galaxies:clusters:A2142--X-rays:galaxies:clusters}
\maketitle

\section{Introduction}
One of the main \chandra\ contributions to the astrophysics of galaxy clusters has been  the discovery of cold fronts. Cold fronts are sharp surface brightness discontinuities, which are clearly visible with high resolution instruments in X-ray images and are interpreted as contact edges between regions of gas with a different entropy (\citealt{mark07}, MV07, for a review). Cold fronts appear almost ubiquitously in galaxy clusters \citep{mark03, ghizzardi10} with the majority of clusters (both relaxed and unrelaxed) hosting at least one cold front.\\
In merging systems (e.\,g.\, Abell 3667, \citealt{vikh01b}), cold fronts were originally interpreted as the discontinuity between the hot shocked ICM and the low-entropy gas of the core of an interacting substructure, which have just survived a pericentric passage. This remnant core interpretation (as dubbed by \citealt{owers09}) was originally proposed by   \citet{mark00} and can be used to explain the origin of cold fronts in many clusters with the most notable example being the ``bullet'' cluster \citep{mark02b}. Soon after this discovery, \chandra\ began to observe cold fronts not only in interacting systems but also in relaxed cool-core clusters (e.\,g.\, A1795 \citealt{mark01a}, RXJ$1720.1+2638$ \citealt{mazzotta01}). These cold fronts are difficult to reconcile with the remnant core interpretation and they are caused by the relative motion of low entropy gas of the cool core in the hotter outer gas. In the now dominant interpretation, this motion, dubbed sloshing by \citet{mark01a},  is a consequence of the perturbation in the gravitational potential of the cluster that follows an off-axis minor merger \citep{tittley05,ascas06}.\\
 The first object in which cold fronts have been revealed by \chandra\ is A2142 (\citealt{mark00}, M00), a hot ($T\sim9$ keV) and X-ray luminous 
cluster at redshift $z=0.091$. The \chandra\ image  shows two prominent surface brightness discontinuities, one at  $\sim 3^{
\prime}$ NW of the cluster center and the other at $\sim 1^{\prime}$ south of the center. Recently, a third surface brightness discontinuity has been identified with a deeper \chandra\ exposure at smaller radii (M.~Markevitch, private communication).
The analysis of the temperature and pressure profile across the discontinuities (M00) has clearly shown that they are cold fronts. Thanks to the striking edges in the \chandra\ image, visible even to an untrained eye, they are often shown as  typical cold front examples,  and therefore, A2142 can be considered the ``archetype'' cold front object. \\
However, the position of A2142 is not well defined in the now dominant classification (MV07) of cold fronts between ``remnant core'' and ``sloshing''. In the paper with the discovery, M00 interpreted the two fronts as remnant cores that survived an off-axis merger in the plane of the sky, but also highlighted the difficulties of this scenario with the existing data on the galaxy distribution. 
A2142 hosts two `` brightest cluster galaxies'' (BCGs): one of them is close to the X-ray peak and therefore probably associated to the southern cold front, while the second one whose position would suggest an association with the NW cold front, has a large peculiar velocity ($v\sim 1800\,\rm{km}\,\rm{s}^{-1}$ \citealt{oegerle95}), making its association with a merging subcluster that moves mainly on the plane of the sky unlikely (M00). \\
When the sloshing mechanism was suggested to explain the presence of cold fronts in relaxed clusters, the striking similarity of the edges in the \chandra\ image of A2142 with the concentric arcs seen in simulation led to the change of the interpretation of cold fronts in this cluster as being caused by the sloshing of the core \citep{tittley05,mark07}. Recently, \citet{owers11} studied the  galaxy distributions in A2142 and, comparing it to the X-ray morphology, identified two candidate perturbers that could have induced the sloshing of the core and that move mainly along the line of sight. In this scenario, the typical sloshing spiral that is coplanar with the perturber orbit is seen edge on, creating the concentric arc features observed in the \chandra\ maps. However, A2142 is very different from the typical clusters where we usually observe sloshing cold fronts. \citet{ascas06} report the presence of a steep entropy profile  as a necessary condition for the onset of the sloshing mechanism. On the observational side, all clusters that host a  sloshing cold front in the sample of \citet{ghizzardi10} feature steep entropy profiles. This is not the case for A2142: its entropy profile is relatively flat, as shown by its central entropy $K_0=68\,\rm{kev \, cm}^2$ \citep{cava09}, which would not classify this object as a typical cool core but as an intermediate object. \\
The available multiwavelength data on A2142 do not provide a clear picture of its dynamical state. The X-ray image is regular, except for the cold fronts, but more elliptical than what usually is found in relaxed clusters.
The \chandra\ temperature map \citep{owers09} shows the presence of cooler ($\sim7$ keV) gas at the center of the cluster and indicates deviations from spherical symmetry at scales $\sim500$ kpc with cooler regions  ($\sim7$ keV) to the SE. 
On the optical side, a deep analysis of spectroscopically confirmed member galaxies by \citet{owers11} highlighted the presence of a number of group-sized substructures consistent with a minor merger. Indications of substructure in the mass distribution come also from the weak lensing data of \citet{okabe08}. On the radio side, \citet{gio00} detected extended emission in the central region of A2142, which was classified as a mini radio halo because of its small size (less than 500 kpc). However, low brightness radio emission extended on Mpc scale has been recently reported for this cluster (Farnsworth et al.~2013, in prep.). This emission is more similar to giant radio halos, which are usually found in merging, non cool core objects (e.\,g.\, \citealt{cassano10, rossetti11}).\\
The cold fronts of A2142 are often considered the textbook examples of cold fronts, yet their host cluster does not fit well within the sloshing paradigm. This makes A2142 an extremely interesting object to test and challenge our understanding of cold fronts and of the sloshing mechanism. For this reason, we asked for and obtained a new \xmm\ observation of A2142, which we present in this paper.\\
Throughout the paper, we adopt $\Lambda$CDM cosmology with $H_{0} = 70$~ km\
s$^{-1}$\ Mpc$^{-1}$, $\Omega_{\rm M} = 0.3$, and $\Omega_\Lambda= 0.7$.
At the redshift of the cluster, $z=0.091$, one arcsecond
corresponds to 1.7 kpc.

\section{Data reduction and analysis}
A2142 has been first observed by \xmm\ in 2002, but the observations were all badly contaminated by soft protons. We obtained a new \xmm\ observation in AO10, and the cluster was observed in July 2011 with an exposure of 55 ks. 

We generated calibrated event files using the SAS software v.\,
$11.0$, and then we removed soft proton flares using a double filtering process: first in a hard ($10-12$ keV) energy band and then in a soft ($2-5$ keV) band.  The observation was not affected by flares, leaving a clean exposure time of more than 50 ks for each detector. The ``in-over-out ratio''  \citep{deluca03}, $R_{SB}<1.1$ for all detectors, indicate only a small contamination from quiescent soft protons. 
The event files were then filtered, according to {\verb PATTERN } and {\verb FLAG } criteria. Bright point-like sources were detected using a procedure based on the SAS task  {\verb edetect_chain } and masked during the analysis. \\

\section{Results}
\begin{figure}
\includegraphics[width=\hsize]{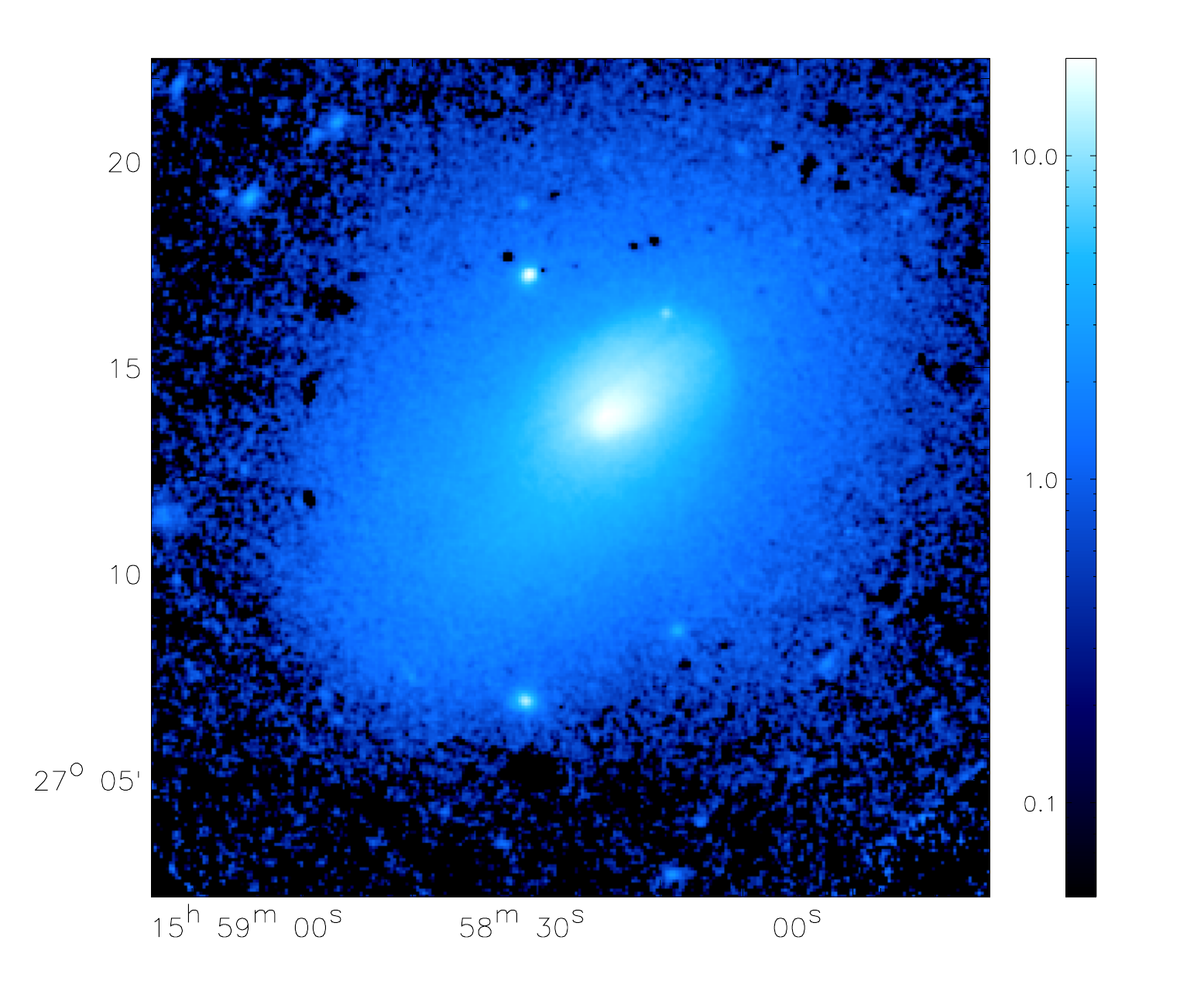}
\caption{EPIC flux image in the energy range $0.4-2$ keV of A2142 in units of $10^{-15}\,\textrm{erg}\,\textrm{cm}^{-2}\,\textrm{s}^{-1}\,\textrm{pixel}^{-1}$ (see \citealt{rossetti06} for details on the preparation of flux images). Coordinates in the image are right ascension and declination.}
\label{fig:sbmap}
\end{figure}
 In Fig.\, \ref{fig:sbmap}, we show an EPIC flux image in the $0.4-2$ keV energy range obtained with the \xmm\ observation of A2142. Despite the lower angular resolution, the two cold fronts discovered by \chandra\ (M00) are also clearly seen in the \xmm\ image as surface brightness edges.
 A third surface brightness edge can be easily recognized in this image about 10 arcmin southeast from the center. 
This newly detected feature was not noticed in earlier studies, since it lies outside of the field of view (FOV) of the existing \emph{Chandra} pointings. Indeed, an excess emission in the SE direction was previously noted by \citet{owers11} and was found to coincide with a tail of cooler gas extending in that direction.\\

\subsection{The SE discontinuity}
\label{sec:se_cf}
\begin{figure*}
\resizebox{\hsize}{!}{\hbox{\includegraphics[height=7cm,width=8.5cm]{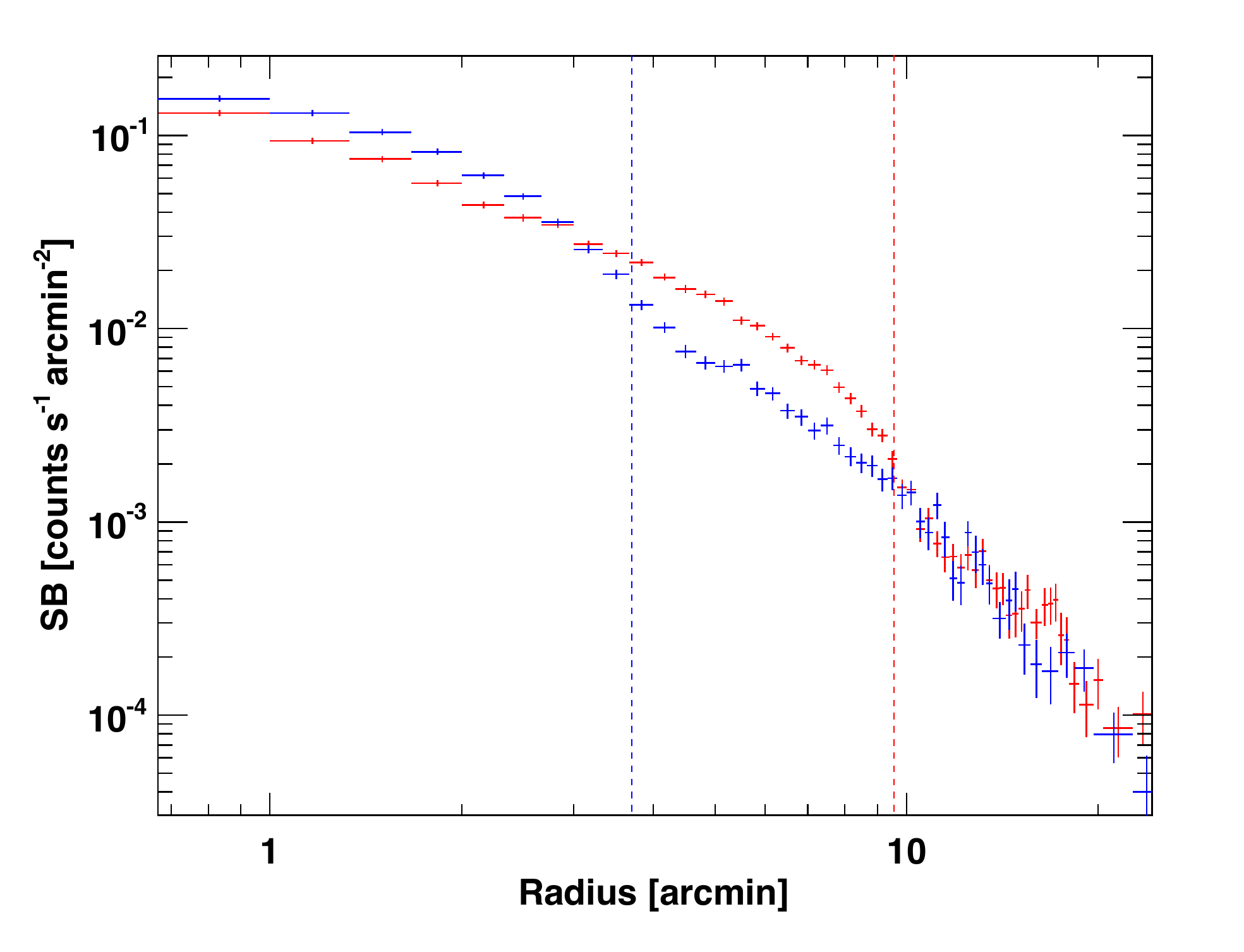}\includegraphics[height=7cm,width=8.5cm]{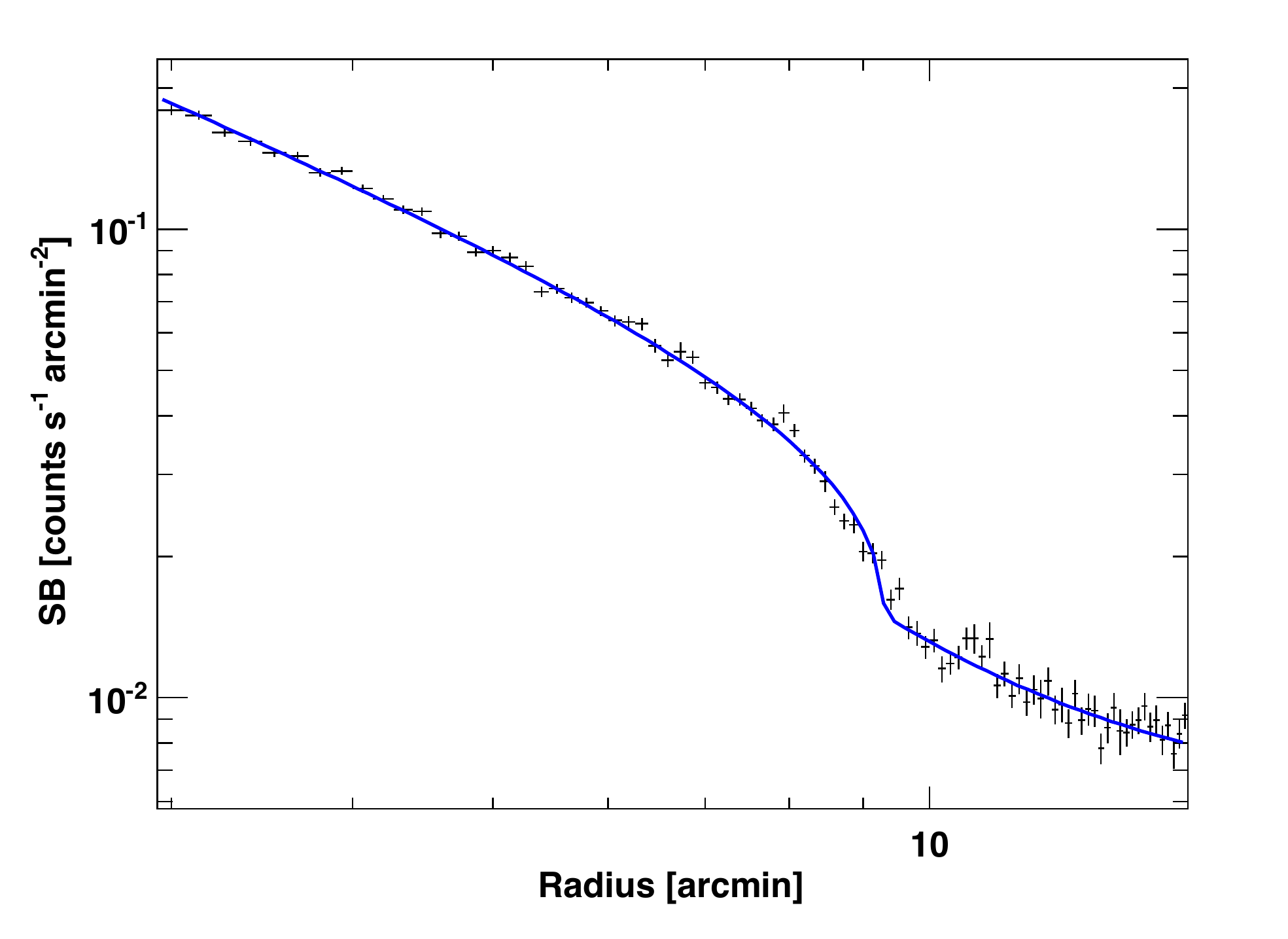}}}
\caption{\emph{Left:} \emph{ROSAT}/PSPC background-subtracted surface-brightness profiles from the surface-brightness peak in the sectors with position angles 0-90$^\circ$ (blue) and 180-270$^\circ$ (red). The dashed lines show the position of the well-known NW cold front and of the newly-discovered SE discontinuity. \emph{Right:} EPIC/pn surface-brightness profile across the SE discontinuity (position angles 180-250$^\circ$) in the 0.5-2.0 keV band, fitted with a broken power law projected along the line of sight (blue curve) and smoothed to account for the \emph{XMM-Newton} PSF.}
\label{fig:se_sbprof}
\end{figure*}
The newly detected feature to the SE was not completely unexpected: after the submission of the \xmm\ AO10 proposal, we noticed in the \rosat/PSPC image of A2142 while analyzing the sample of clusters discussed in \citet{eckert12}. While \rosat/PSPC data allowed us to detect the surface brightness discontinuity in the profile (Sec.~\ref{sec:se_ima}), the \xmm\ data were needed to spectroscopically assess the nature of this new feature (Sec.~\ref{sec:se_spe}). 
\subsubsection{Imaging analysis}
\label{sec:se_ima}
We analyzed the \rosat\ PSPC pointed observation of A2142, as discussed in \citet{eckert12}, and prepared an exposure-corrected background-subtracted\footnote{The background image that we subtracted from the source includes the non-cosmic components (the particle background, the scattered solar X-rays and the long-term enhancement) as modeled by the ESAS package \citep{snowden94} } image in the R37 band ($0.42-2.01$ keV). We extracted the PSPC surface-brightness profile from the surface-brightness peak in a sector with position angles 180-270$^\circ$ (all angles are measured counterclockwise from the right ascension axis). We used elliptical annuli to follow the X-ray isophotes as closely as possible. The resulting profile is shown in the left hand panel of Fig. \ref{fig:se_sbprof} and compared to the counter-sector (position angles 0-90$^\circ$). The SE profile shows a clear excess in the 4-8$^\prime$ range compared to the azimuthal average, followed by an abrupt surface-brightness drop, that is indicative of a density discontinuity. Beyond $\sim10^{\prime}$, the behavior of the SE sector again agrees with the azimuthal average.\\
We confirmed the discontinuity to the SE with the higher resolution \xmm\ data: the pn profile is shown in 
the right panel of Fig.~\ref{fig:se_sbprof}. . 
We fitted it with a projected broken power law elliptical model (e.~g.~\citealt{owers09}), convolved with the point spread function (PSF) of the instrument.  The slope of the outer density profile was fixed at the value derived from \emph{ROSAT} data, since the \xmm\ profile does not extend much beyond the discontinuity. The best-fitting profile is overplotted in Fig. \ref{fig:se_sbprof},  and the parameters are shown in Table \ref{tab:se_cf}. The density jump is well constrained by the pn data, where $n_{in}/n_{out}=1.79_{-0.05}^{+0.06}$, and consistent results are found with the MOS profiles. The model gives a distance $r_{cut}=947 \pm 3$ kpc for the density discontinuity from the surface-brightness peak, which corresponds roughly to half of the virial radius of the system.\\
Although the shape of the profile (Fig.~\ref{fig:se_sbprof}) is very typical of surface brightness discontinuities and the broken power-law elliptical model provides a good fit to the data, we also fitted it with other models to confirm the presence of a discontinuity. Details of the analysis are provided in Appendix \ref{app:sbprof}: simple models (single or double power-law) cannot reproduce the observed data. A continuous triple power-law provides a slightly improved fit to the data but requires the middle slope to be very steep ($\alpha \simeq -6$), which is very similar to the broken power law case. From this analysis, we can safely confirm the presence of a discontinuity in the profile shown in Fig. ~\ref{fig:se_sbprof}.\\
\begin{table}
\caption{\label{tab:se_cf}Best-fitting parameters of the broken power law model to the EPIC-pn data. Errors are quoted at the $1\sigma$ level.}
\begin{center}
\begin{tabular}{cc}
\hline
Parameter & Value \\
\hline
\hline
$\alpha_{in}$ & $1.12\pm0.01$\\
$\alpha_{out}$ & $2.05\pm0.08$\\
$r_{cut}$ & $947\pm3$\\
$n_{in}/n_{out}$ & $1.79_{-0.05}^{+0.06}$\\
\hline
\end{tabular}
\end{center}
\textbf{Notes:} Parameter description: $\alpha_{in,out}$ are the slopes of the inner and outer density profiles, $r_{cut}$ is the break radius in kpc, and $n_{in}/n_{out}$ is the  density jump across the front.
\end{table}
We investigated the aperture of the front by extracting surface-brightness profiles in narrow sectors of $10^\circ$ opening. In each sector, we applied the criterium described in \citet{ghizzardi10} to assess the presence of a surface-brightness discontinuity. Namely, we fitted the profiles with a power law in the radial ranges corresponding to the front ($7.5^\prime-9.5^\prime$) and  inside ($5^\prime-7.5^\prime$). We declared the detection of a discontinuity, if the difference of slope in these two radial ranges significantly exceeds $0.4$. As a result, we found that a surface-brightness discontinuity is detected in the sectors with position angles $180-250^\circ$, which corresponds to the impressive linear scale of 1.2 Mpc. 

\subsubsection{Spectral analysis}
\label{sec:se_spe}
\begin{figure*}
\resizebox{\hsize}{!}{\hbox{\includegraphics[width=9cm]{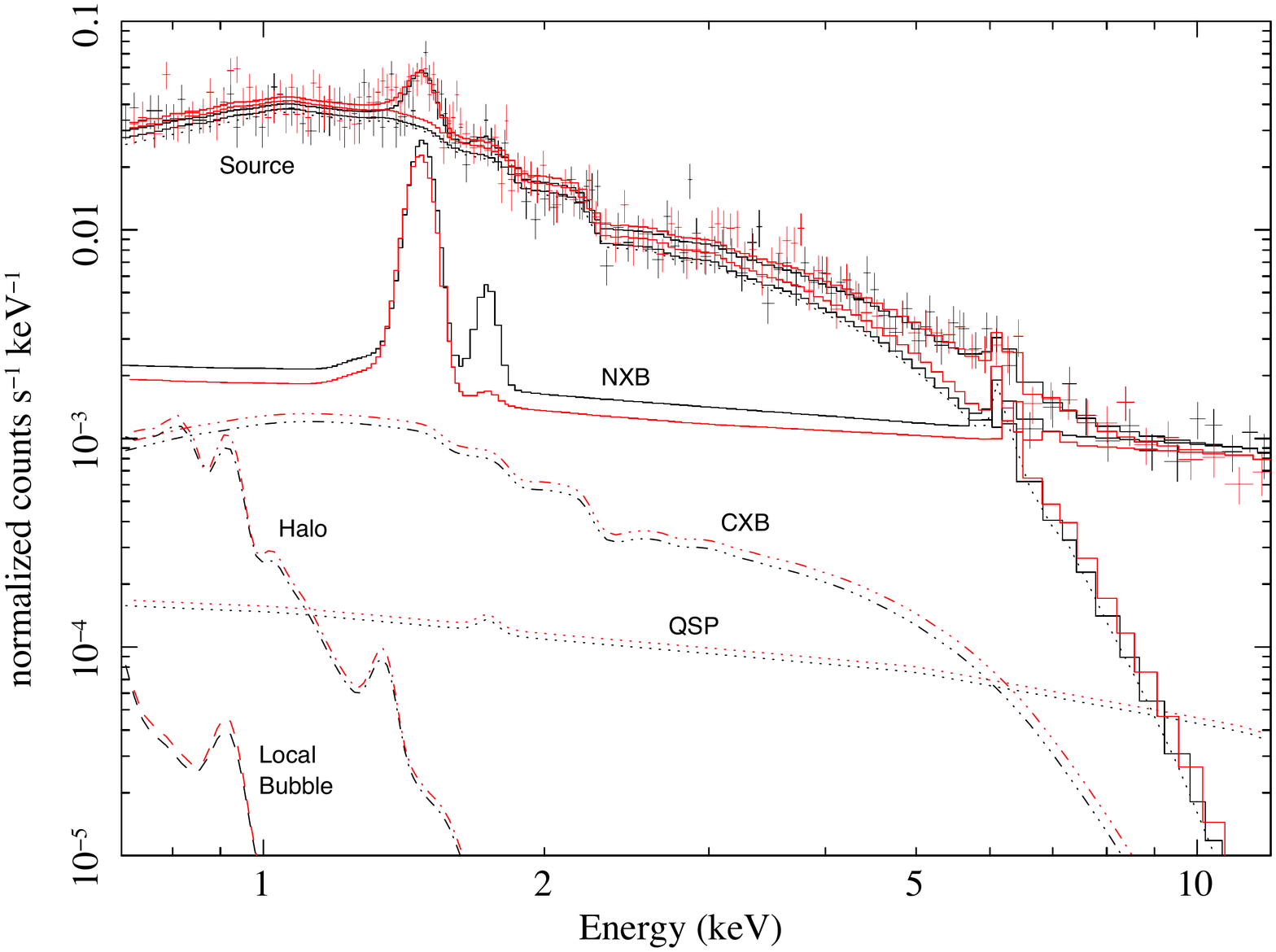}\includegraphics[width=9cm]{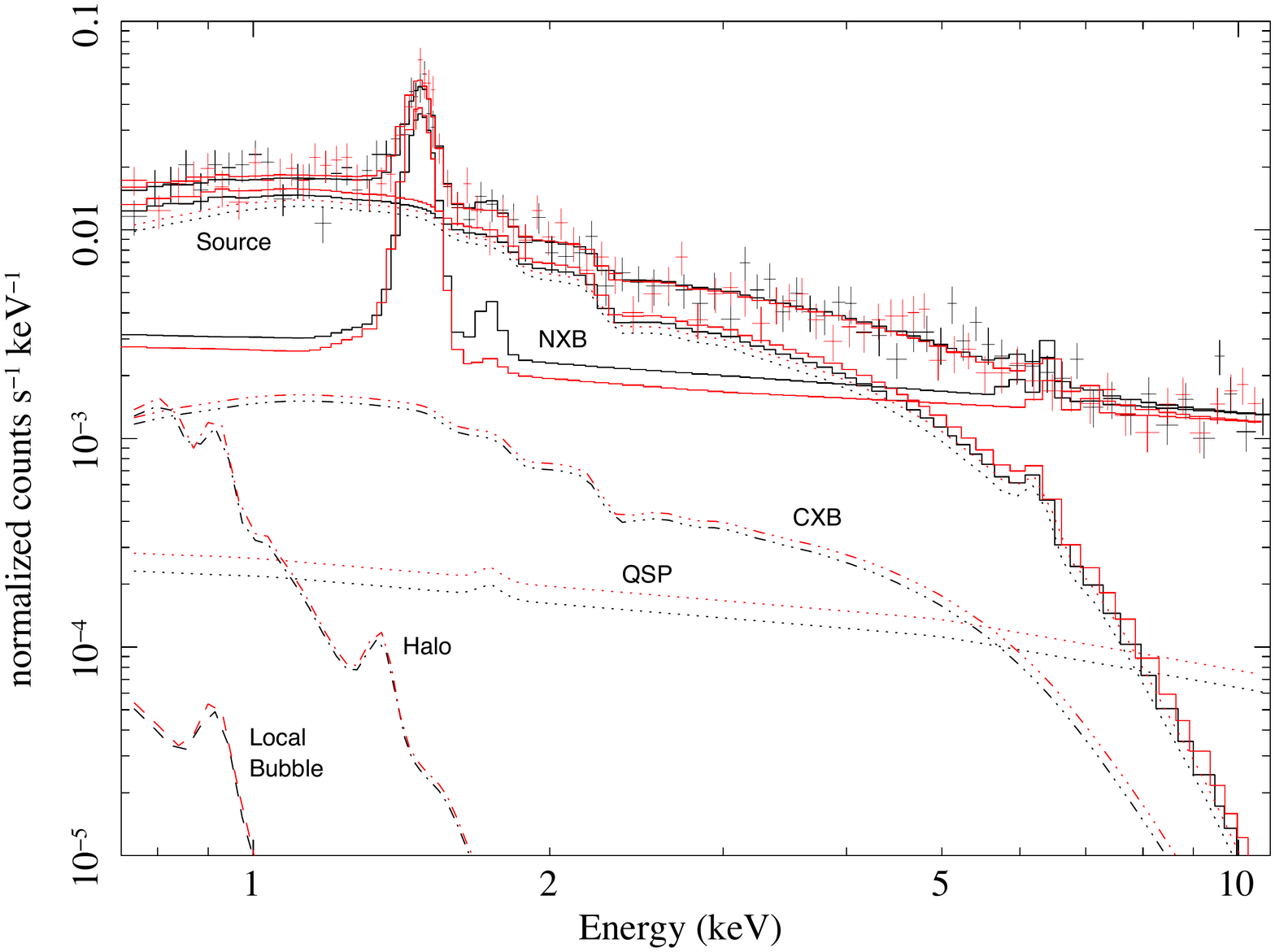}}}
\caption{\emph{XMM-Newton}/MOS spectra of the IN (left) and OUT (right) regions and best-fit models. MOS1 data are shown in black and MOS2 in red. The contribution from the various background components (sky: local bubble, Galactic halo, and CXB; internal: NXB and QSP) is compared to the best-fit source model.}
\label{fig:spec_inout}
\end{figure*}
\begin{figure*}
\resizebox{\hsize}{!}{\hbox{\includegraphics[width=9cm]{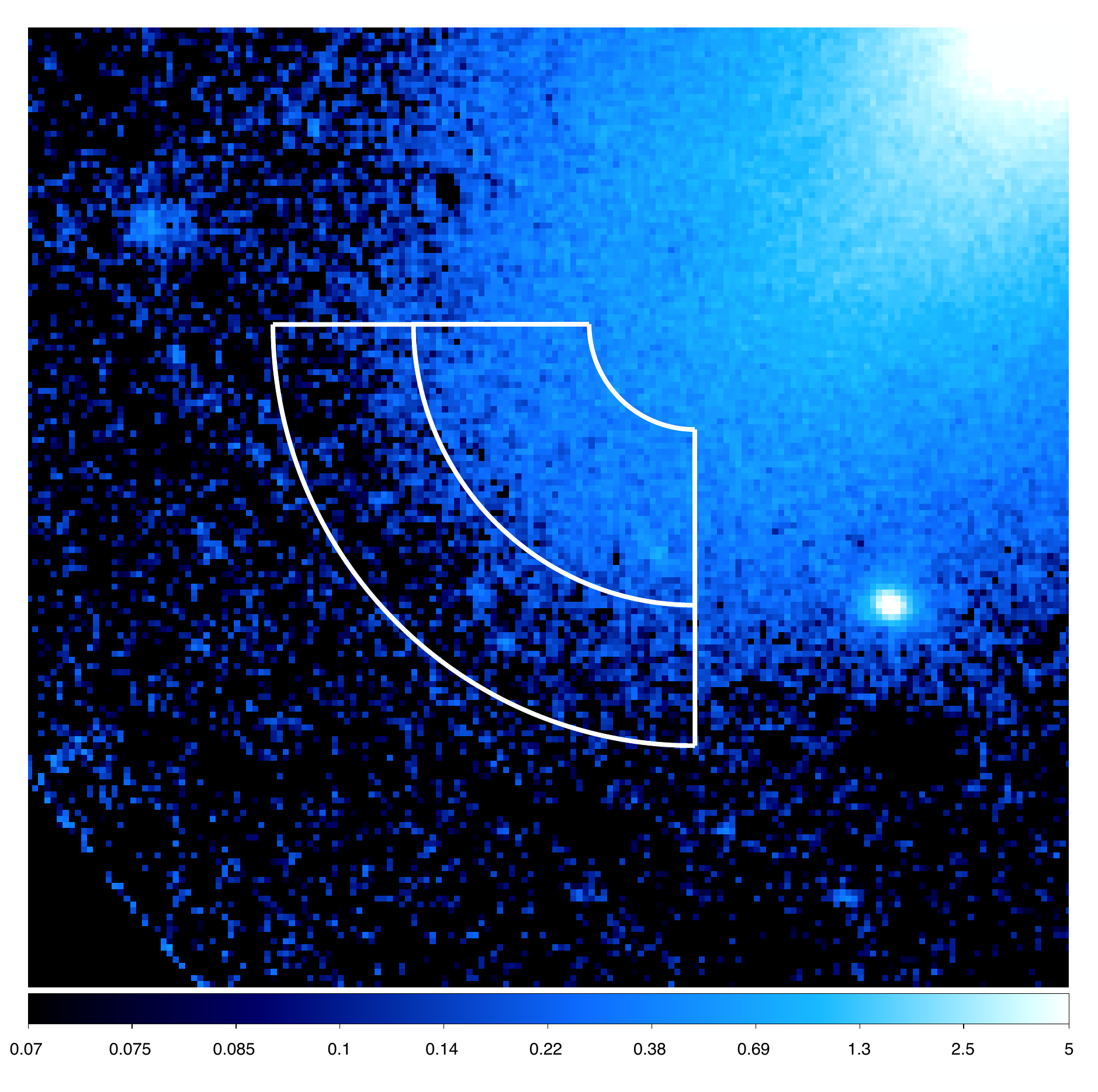}\includegraphics[width=9cm]{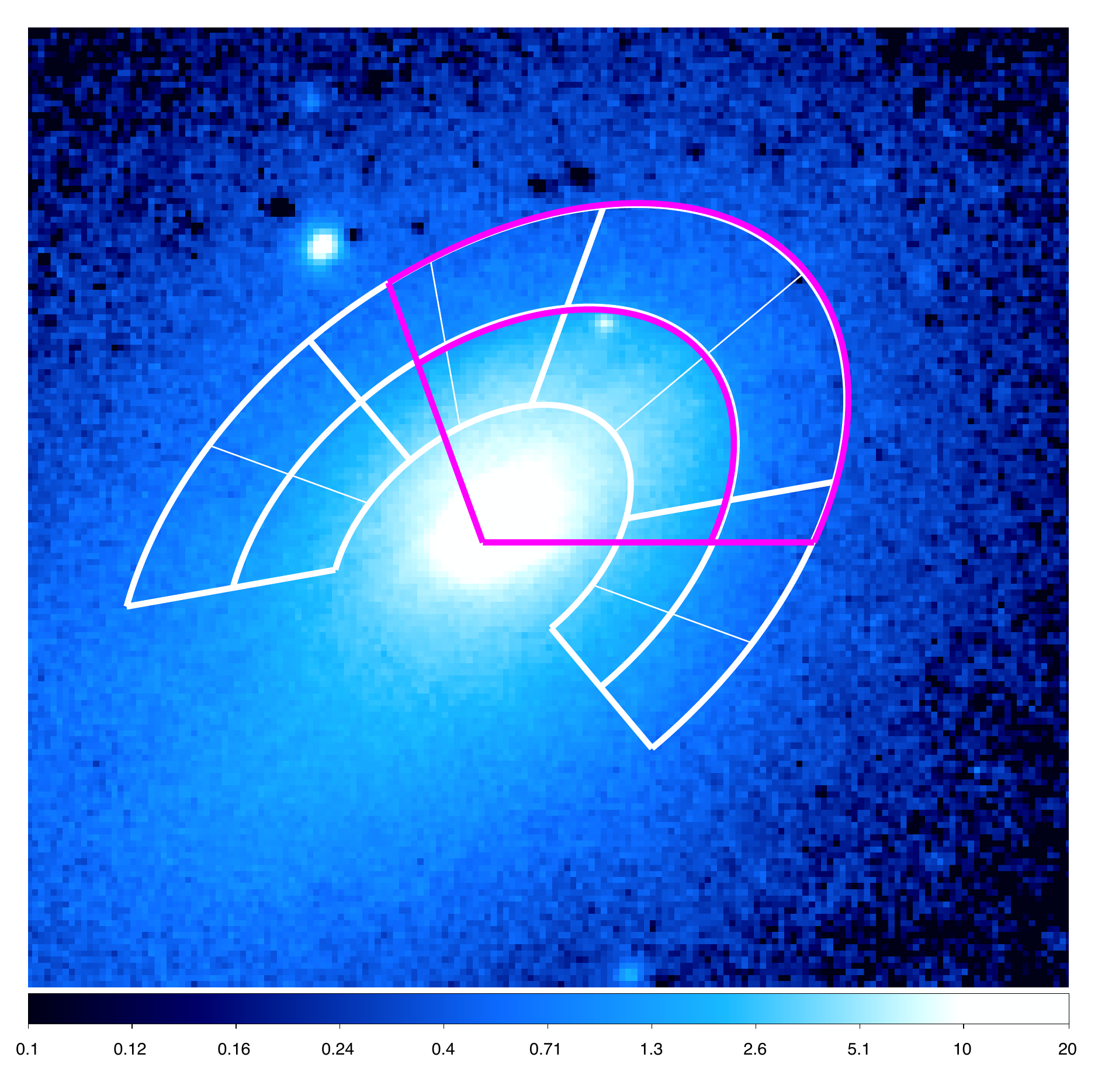}}}
\caption{Zoom on the SE (left) and NW (right) cold fronts with the regions used for spectral and imaging analysis. The magenta region in the right panel has been used for deriving the global properties of the NW cold front (Sec.~\ref{sec:propCF}), the white region for the azimuthal extent  surface brightness (thin lines) and temperature (thick lines) analysis (Sec.~\ref{res_stability}).
The IN and OUT sector for the SE cold front are centered at (RA,Dec)$=(15:58:40,+27:10:12.6)$; the elliptical sectors used for the NW cold front are centered at  (RA,Dec)$=(15:58:21,+27:13:41)$. }
\label{fig:show_reg}
\end{figure*}
The spectral analysis is a necessary step to assess the nature of this newly detected discontinuity. Indeed, surface brightness edges are observed both in cold fronts and in shocks: the sign of the temperature discontinuity is a key information to distinguish between the two scenarios.\\
We extracted spectra in two elliptical sectors across the surface brightness edge (Fig.~\ref{fig:show_reg}, left), generating a response file and an ancillary file with the SAS task {\verb rmfgen } and {\verb arfgen } in extended source mode.  Because of the low surface-brightness of the signal in the OUT region, an improper subtraction of the background could have a strong influence on the result, so a detailed modeling of the various background components is necessary. We applied a method similar to the one described in \citet{leccardi08a} to model the background in the two regions and restricted our analysis to the MOS detectors, since their background was found to be more predictable than for the pn. The details of the background modeling are described in Appendix B.
Spectral fitting to the spectra in the IN and OUT regions was performed using XSPEC v12.7.0 and the modified C-statistic. The appropriate effective area was applied to the sky components, but not to the instrumental ones. The source spectrum was modeled with an absorbed APEC model with metal abundance fixed to $0.3Z_\odot$ and redshift fixed to that of the cluster. All the free parameters were then fitted simultaneously. The resulting spectra are shown in Fig. \ref{fig:spec_inout} with the various model components. In the IN region, our model gives a best-fit temperature of $kT_{IN}=6.13_{-0.37}^{+0.45}$ keV, while we find $kT_{OUT}=9.0_{-1.5}^{+2.4}$ keV  for the OUT region. The errors are given at the $1\sigma$ confidence level.\\ 
The temperature of the OUT region appears higher than that of the IN region, although the temperature difference is only moderately significant, which suggests that the SE discontinuity is likely a cold front. In the shock scenario with an inner temperature of 6.1 keV and a density jump of 1.8 (see Table \ref{tab:se_cf}), the Rankine-Hugoniot conditions predict an outer temperature of 3.9 keV. Fixing the temperature of the plasma to 3.9 keV, we find a difference in C-statistic $\Delta C=28.1$ compared to the fit with $kT_{OUT}=9.0$ keV, which translates into a rejection of this hypothesis at $5.3\sigma$. Therefore, we can conclude with good confidence that this discontinuity is a cold front. At almost 1 Mpc from the center, this is the most distant cold front ever detected in a galaxy cluster (see Sec.~4).

\subsection{The NW cold front}
\label{sec:propCF}
The spatial resolution of \xmm\ is sufficient to analyze the surface brightness and temperature profile across the NW cold front and compare them to results obtained with \chandra. \\
We extracted the surface brightness profile across the NW cold front in the elliptical sector shown in Fig. ~\ref{fig:show_reg} (right panel, magenta sector). 
We fitted the profile between 1 and 7 arcminutes from the center with an elliptical broken power-law model \citep{owers09}, and we found a density jump $n_{IN}/n_{OUT}=1.97\pm0.03$ with the MOS image and $n_{IN}/n_{OUT}=2.00\pm0.04$ with the pn. These values are consistent with the measurement by \citet{owers09}, where $n_{IN}/n_{OUT}=2.0\pm0.1$ , obtained with higher resolution \chandra\ data. We extracted spectra in elliptical annular sectors to measure temperature inside and outside the edge and found $kT_{IN}=(6.9 \pm 0.1)$ keV and  $kT_{OUT}=(10.0 \pm 0.3)$ keV, which is a significant temperature discontinuity. With this analysis, we could also measure a mildly significant ($1.8 \sigma$) metal abundance discontinuity across the front: $Z_{IN}=(0.32 \pm 0.02)Z_\odot$ and $Z_{OUT}=(0.24 \pm 0.04)Z_\odot$. \\
Since the cold front is detected over a large angular scale (see Sec.\,\ref{res_stability}), we chose a sector extending from $-20$ to $190$ degrees, which corresponds to the whole angular range where a density discontinuity is detected. The best fit density jump is  $n_{IN}/n_{OUT}=1.76\pm0.02$. The temperature discontinuity is lower but still significant: $kT_{IN}=(7.42 \pm 0.08)$ keV, and  $kT_{OUT}=(9.6 \pm 0.2)$ keV.\\

\subsubsection{Azimuthal extent of the cold front}
\label{res_stability}
\begin{figure}
\includegraphics[width=\hsize]{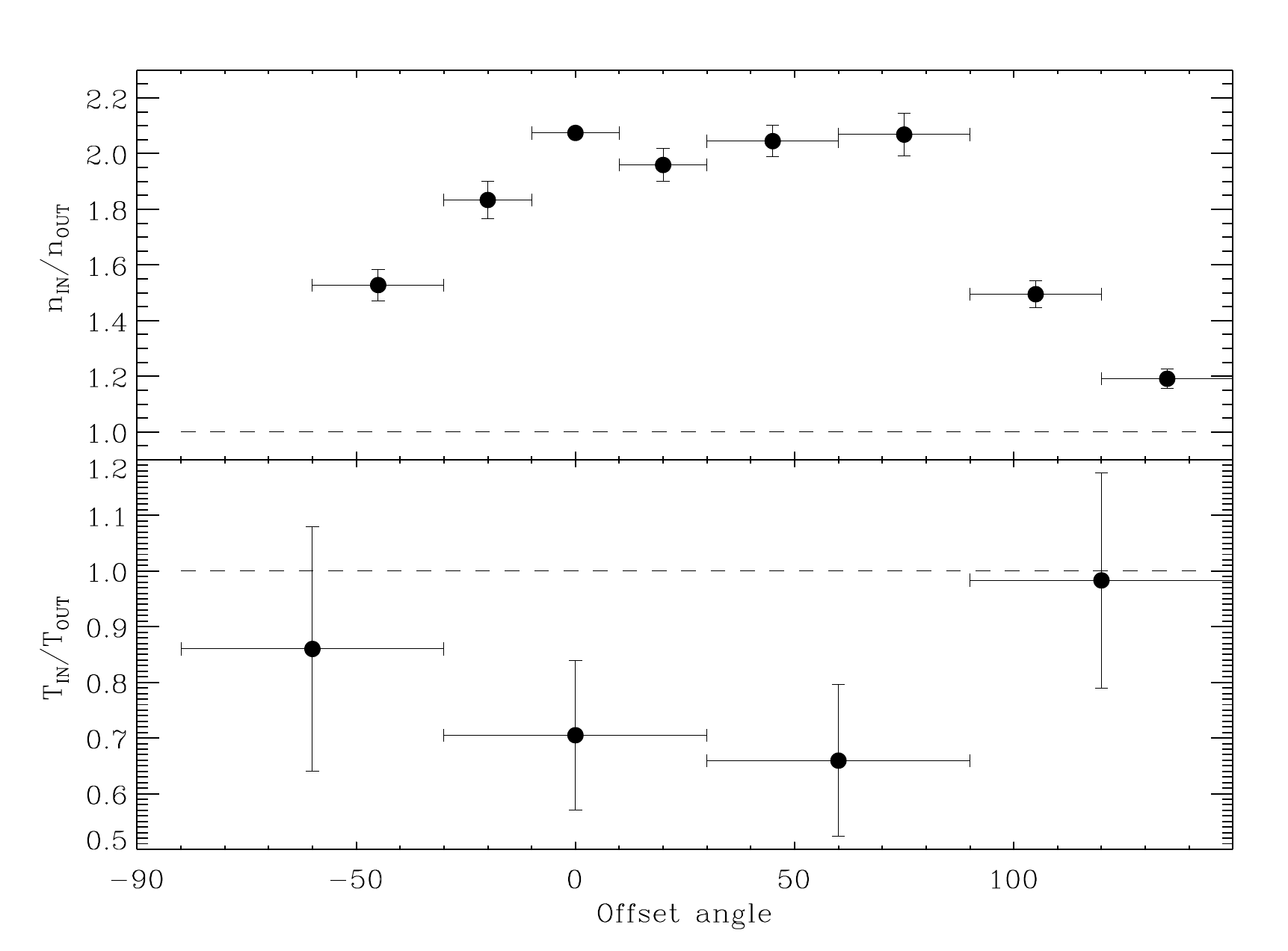}
\caption{Density (upper panel) and temperature (lower panel discontinuities as a function of the off axis angle from the axis inclination of the NW cold front).}
\label{fig:stability}
\end{figure}

The northwestern cold front in A2142 looks sharp over a broad angular range. We extracted the surface brightness profile in narrow (20-30\deg, Fig.~\ref{fig:show_reg}, right panel) elliptical sectors and fitted them with the projected broken power law model, as described above. A significant density discontinuity ($n_{IN}/n_{OUT}>1$ at more than $3\sigma$) is detected from the sector [-20\deg,10\deg] to the sector [160\deg ,190\deg ].
In Fig.~\ref{fig:stability}, we plot the density jump as a function of the off-axis angle from the ``tip'' of the front, that we assume to be 40\deg, coincident with the overall inclination of the elliptical contours. A similar analysis has been performed for the remnant-core.
 cold front of A3667 by \citet{vikh01b}. In A3667 the shape of the front is symmetric and less extended than in A2142, where the discontinuity extends more to the north than to the south. 
The very broad angular range (180\deg) over which the density discontinuity is detected, can provide information on the geometry of the fronts with respect to the line of sight (Sec.~4). \\
We also measured the temperature discontinuity across the cold front in sectors of 60\deg width. The temperature jump is significant at more than 2$\sigma$ in the sector extending from -90 to 130 degrees from the R.A. axis; at larger angles, the discontinuity is not significant.\\
This analysis confirms the impression from the image in Fig.~\ref{fig:show_reg} that the NW cold front remains sharp and smooth over a broad angular range without any clear distortion, at the \xmm\ resolution scale, which may be produced by the onset of hydrodynamical instabilities. 

\subsection{Residual image}
\begin{figure*}
\includegraphics[width=\hsize]{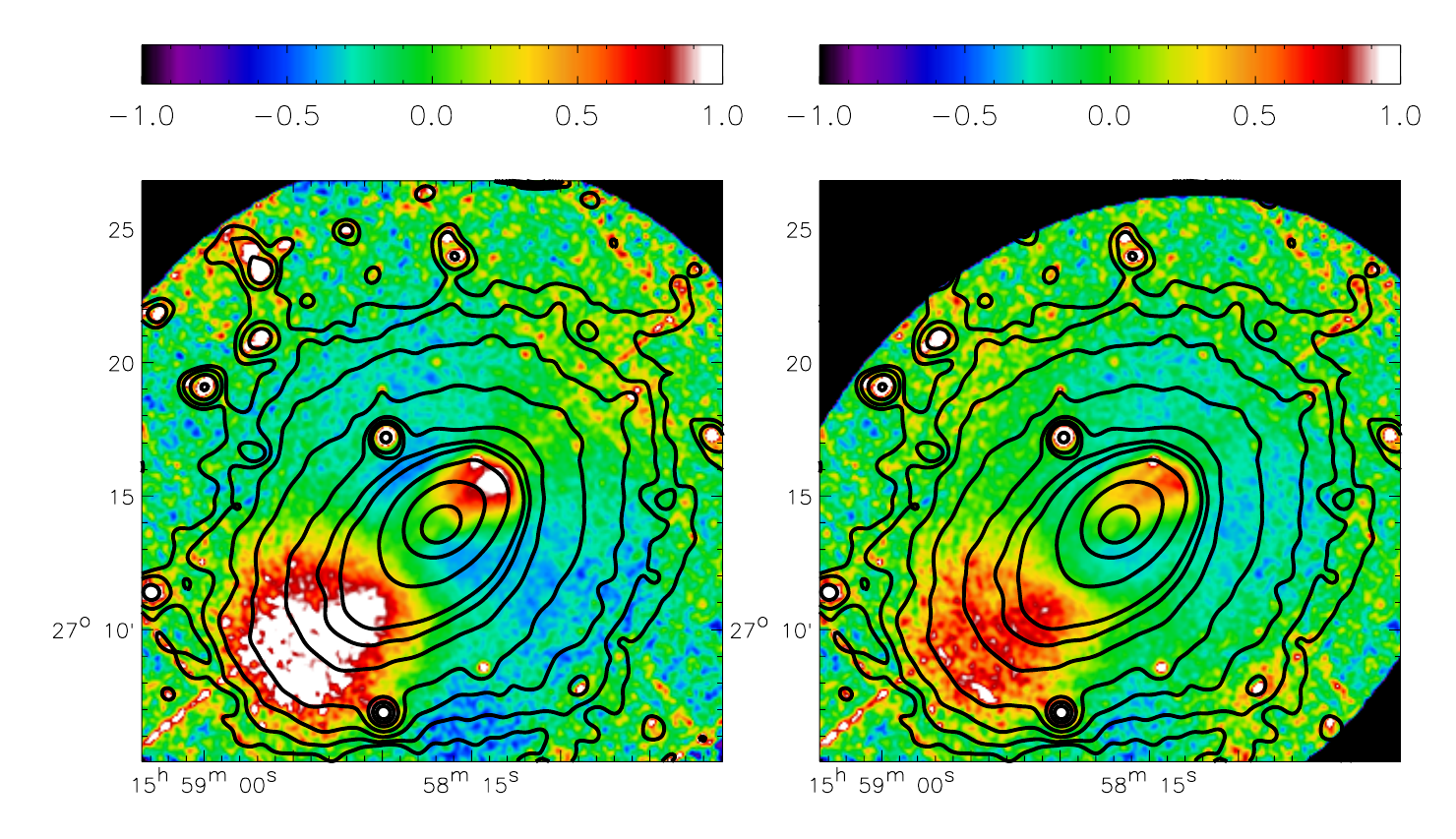}
\caption{Residual image from the azimuthal average in concentric annuli ({\it left}) and in elliptical annuli ({\it right}). X-ray contours are overlaid, coordinates on the images are right ascension and declination.}
\label{fig:residuals}
\end{figure*}

Residual images highlight the  surface brightness deviations from a symmetric model and thus  are an effective way to detect and identify cold fronts, in both real and simulated images \citep{roediger12}.
We prepare residual images starting from the EPIC flux image: we mask point sources and calculate the average surface brightness in concentric annuli (circular or elliptical). We then subtract this symmetric model from the image and divide the resulting difference image by the model. This method allows us to provide quantitative information on the relative variation of the source with respect to the symmetric model. It is less model-dependent than calculating residuals using a beta-model fit of the azimuthal surface brightness profile. \\
The residual map obtained this way is shown in Fig. \ref{fig:residuals}, middle panel. Positive residuals are apparent at the position of the two known cold fronts and a significant large scale excess is clearly detected to the SE, corresponding to the new feature that we discussed in Sec. \ref{sec:se_cf}. Part of this excess was indeed visible also in the \chandra\ residual image shown by \citet{owers11}.\\
Since the overall geometry of the X-ray emission of A2142 is very elliptical (with an axes ratio of $1.58$), we also extracted the surface brightness profile in elliptical sectors and produced the residual image with  this model  (Fig.\, \ref{fig:residuals}, right panel). While the negative residuals in the NE-SW direction almost disappear, confirming that they are only the signature of the ellipticity, the concentric excesses corresponding to the three cold fronts are still clearly visible, showing that they are robust against the assumed geometry.\\
The presence of multiple concentric excesses or spiral-like features in the residual map is often observed in residual images of simulated clusters featuring sloshing (see Sec.~\ref{sec:discussion} for more details).

\subsection{Thermodynamic maps and X-ray morphology}
\label{sec:maps_morph}
\begin{figure*}
\includegraphics[width=\hsize]{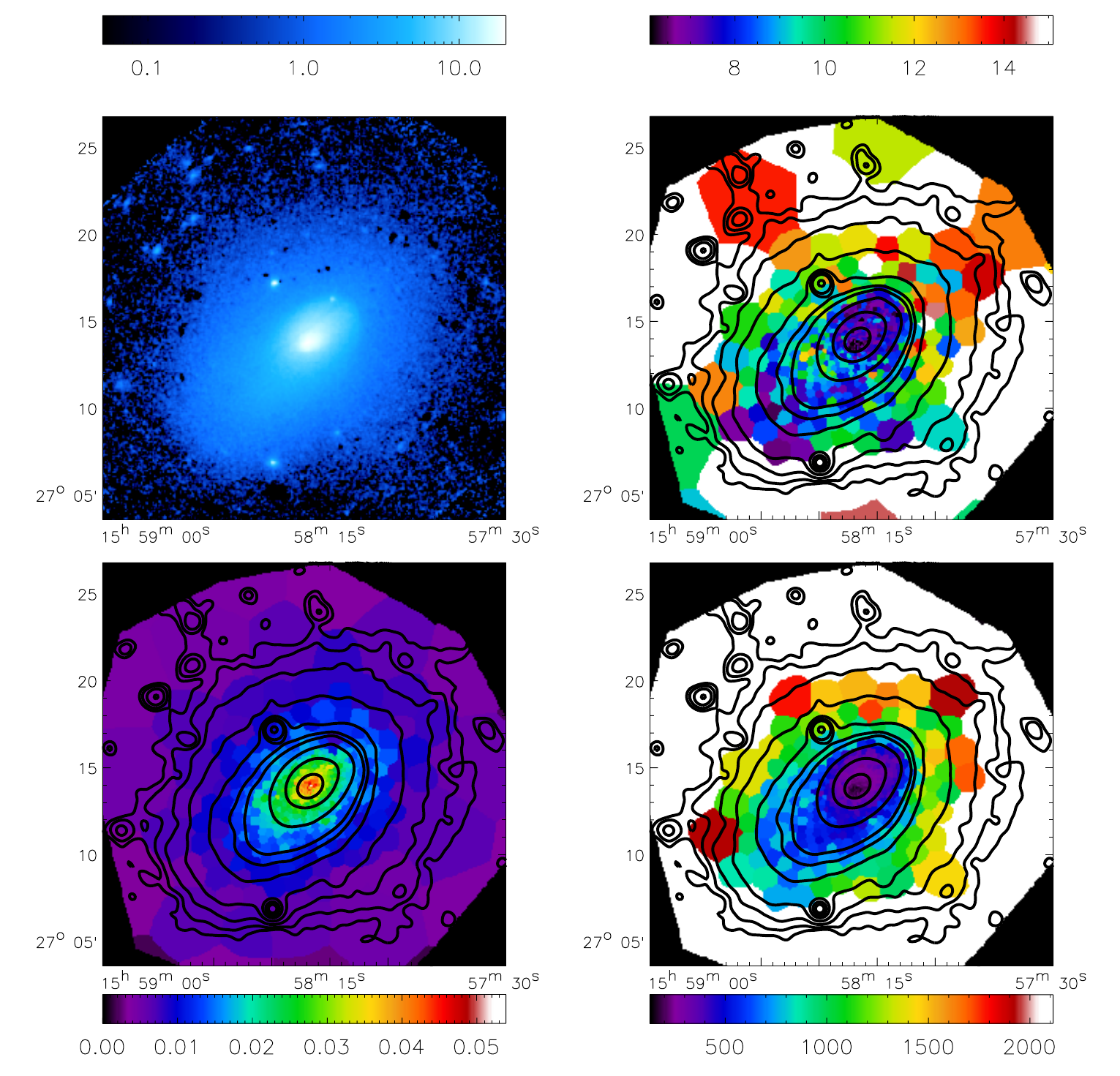}
\caption{EPIC SB image (upper left), temperature map(upper right), pressure map(lower left) and entropy map (lower right). Maps were obtained with a target $S/N=15$. X-ray contours are overlaid on the thermodynamic maps, coordinates are right ascension and declination. }
\label{fig:mappe}
\end{figure*}
We used the technique described in \citet{rossetti06} to prepare two--dimensional temperature and projected density maps starting from X-ray images. We then combined them to produce pseudo-pressure and pseudo-entropy maps.
Our maps, shown in Fig.\,\ref{fig:mappe}, can be used to infer the gas properties in the more central region, up to the SE cold front but do not allow us to study the outskirts, since they were obtained with a simple background subtraction.
The temperature map shows the presence of cooler gas (around 7 keV) in the denser regions of all three cold fronts detected in the X-ray image. The pseudo-entropy map shows that the lowest entropy gas is located close to the more central SE cold front, as observed in sloshing simulation, and shows an extension of lower entropy gas to the SE in the direction of the new cold front. Both maps are highly asymmetric at all scales, as usually observed in non cool core unrelaxed clusters. The pseudo pressure map is more symmetric but elliptical, possibly reflecting the shape of the underlying potential well of the cluster.\\
To better characterize the dynamical state of A2142, we also measured different morphological estimators starting from X-ray images (Table \ref{tab:morpho}). We calculated the centroid shift $w$, within both $R_{500}$ \citep{maughan08} and 500 kpc \citep{cassano10}: in both cases, the values are at the limit between the relaxed and disturbed class. The value of the surface brightness concentration parameter $c$ depends on the scales used to quantify it: at the intermediate scales mapped with the definition in \citep{cassano10}, $c_{100}$=SB($<100$ kpc)/SB($<500$ kpc) the cluster appears relaxed, while at the smaller scales mapped by the original definition of \citep{santos08}, $c_{40}=$SB($<40$ kpc)/SB($<400$ kpc), A2142 is not as concentrated as cool core objects. The ellipticity is very high: compared to the sample of \citet{hashimoto07}, A2142 shows one of the largest values ($\epsilon=1-b/a=0.37$). \\
We can also study the thermodynamical state of A2142 using indicators of the cool core state. The central entropy ($K_0=68\pm2$ keV cm$^2$) is higher than the typical values of cool core objects \citep{cava09}, and the pseudo-entropy ratio classifies it as an intermediate objects \citep{leccardi10}. \\
According to these results (summarized in Table \ref{tab:morpho}), A2142 cannot be unambiguously classified either as a relaxed cool core object or as an unrelaxed merging non cool core cluster. Its characteristics are intermediate between the two classes. 
\begin{table}
\caption{Morphological and thermodynamical indicators of A2142}
\label{tab:morpho}
\centering
\begin{tabular}{c c c c}
\hline \hline
Indicator & value & class & references\\
\hline
$w(<R_{500})$ & $(1.4\pm0.1)\times10^{-2}$ & intermediate & 1 \\
$w(<500\,\rm{kpc})$ & $(1.3\pm0.1)\times10^{-2}$ & intermediate & 2 \\
$c_{100}$ & $0.247\pm0.001$ & relaxed & 2 \\
$c_{40}$ & $0.069\pm0.001$ & non-cool core & 3 \\
$\epsilon$ & $0.37 \pm 0.01$ & unrelaxed & 4 \\
$K_0$ & $(68\pm2)\rm{keV\,cm^2}$ & non-cool core & 5 \\
$\sigma$ & $(0.48 \pm 0.01)$ & intermediate & 6 \\
\hline
\end{tabular}
\tablebib{
(1)~\citet{maughan08}; (2)~\citet{cassano10}; (3)~\citet{santos08}; (4)~\citet{hashimoto07}; (5)~\citet{cava09}; (6) \citet{leccardi10}.
}
\end{table}

\subsection{Metal abundance distribution}
\label{sec:metal}
We selected regions for spectral analysis starting from the surface brightness residual map. This selection allows us to characterize the thermodynamic properties of the ICM featuring a surface brightness excess compared to the other regions of the cluster. More specifically, if the SB excess is caused by the ICM of the core ``sloshing'' around the cluster, we expect these regions to be cooler and possibly metal richer compared to other regions at the same distance from the cluster center. We selected manually the regions in the residual surface brightness map,  and we extracted spectra (as in Sec.~\ref{sec:se_spe}) from each region. We then fitted spectra within XSPEC in the $0.7-10$ keV band with a {\it vapec} model multiplied by the Galactic hydrogen column density, $N_H$,  through the {\it wabs} absorption model. In the fitting procedure, we fixed redshift and $N_H$, whereas we left as free parameters the abundances of all the elements with emission lines at energies above $0.7$ keV. The metal abundances are measured relative to the solar photospheric values of \citet{anders89}. 
Regions corresponding to the SB excess associated to the central and northwestern cold fronts  feature lower temperature and higher metal abundance, $Z=(0.35\pm0.1)Z_\odot$,  than other regions located at the same distance from the center, $Z=(0.29\pm0.1) Z_\odot$. Although the metal abundance excess is small, it is interesting to show that the gas with the highest metallicity (and lower temperature) of the cluster is located in the regions corresponding to the central and NW cold fronts. The regions corresponding to the SE large scale excess and to the third cold front also feature a lower temperature with respect to regions at the same distance from the center but do not show a significant metal abundance excess. This is expected since sloshing does not displace gas by more than a factor of 2-3 in terms of distance from the center \citep{ascas06}, therefore gas in the SE excess likely originates in a region not significantly enriched.\\
Following \citet{rossetti10}, the regions featuring a surface brightness excess along with lower temperature and high metal abundance can be classified as a cool core remnant. This means that A2142 must have undergone an earlier cool core phase in its evolution, whose signature was later destroyed by a merger.

\section{Discussion}
\label{sec:discussion}
\subsection{The SE cold front}
As discussed in Sec.~\ref{sec:se_cf}, the clearly visible southeastern edge in Fig.~1 is likely a cold front. The presence of another cold front in this cluster, located at almost 1 Mpc from the center, makes the overall picture described in Sec.~1 even more intriguing. \\
We tested the possibility that this cold front may belong to the remnant-core class and therefore be the boundary of a merging subcluster traveling toward southeast in the ICM of the main cluster. This merger could possibly induce the sloshing of the core to produce the other small scale cold fronts. However, we could not find either a galaxy concentration in the analysis by \citet{owers11} or a clump in the mass distribution in the weak-lensing analysis \citep{okabe08}, which could be associated to this feature.\\
The presence of this newly detected discontinuity cannot be understood without a global view of the cluster. A2142 is one of the few clusters hosting four cold fronts, with A496 \citep{ghizzardi10} and possibly Perseus \citep{simio12}. 
 Moreover, the large scale excess looks connected to the smaller scale structures: the SE excess begins, in terms of radius, at the same distance from the center as the NW discontinuity (see Fig. 2, left panel). This alternation of excesses and crossing of profiles in opposite directions is a generic feature of the sloshing scenario \citep{roediger12} and was noticed also in Perseus by \citet{simio12}.\\
The SE cold front of A2142 is a record breaking feature: at 1 Mpc from the center, it is the outermost cold front, detected as both a surface brightness and temperature discontinuity and not obviously associated to a moving subcluster, ever observed in a galaxy cluster (see also Sec.~\ref{sec:compare}). Moreover, it is detectable as a surface brightness discontinuity over a broad angular range (70 degrees), which corresponds to a linear scale of 1.2 Mpc at a distance from the center of 1 Mpc.

\subsection{Comparison with simulations}
\label{sec:simul}
\begin{figure}
\includegraphics[width=\hsize]{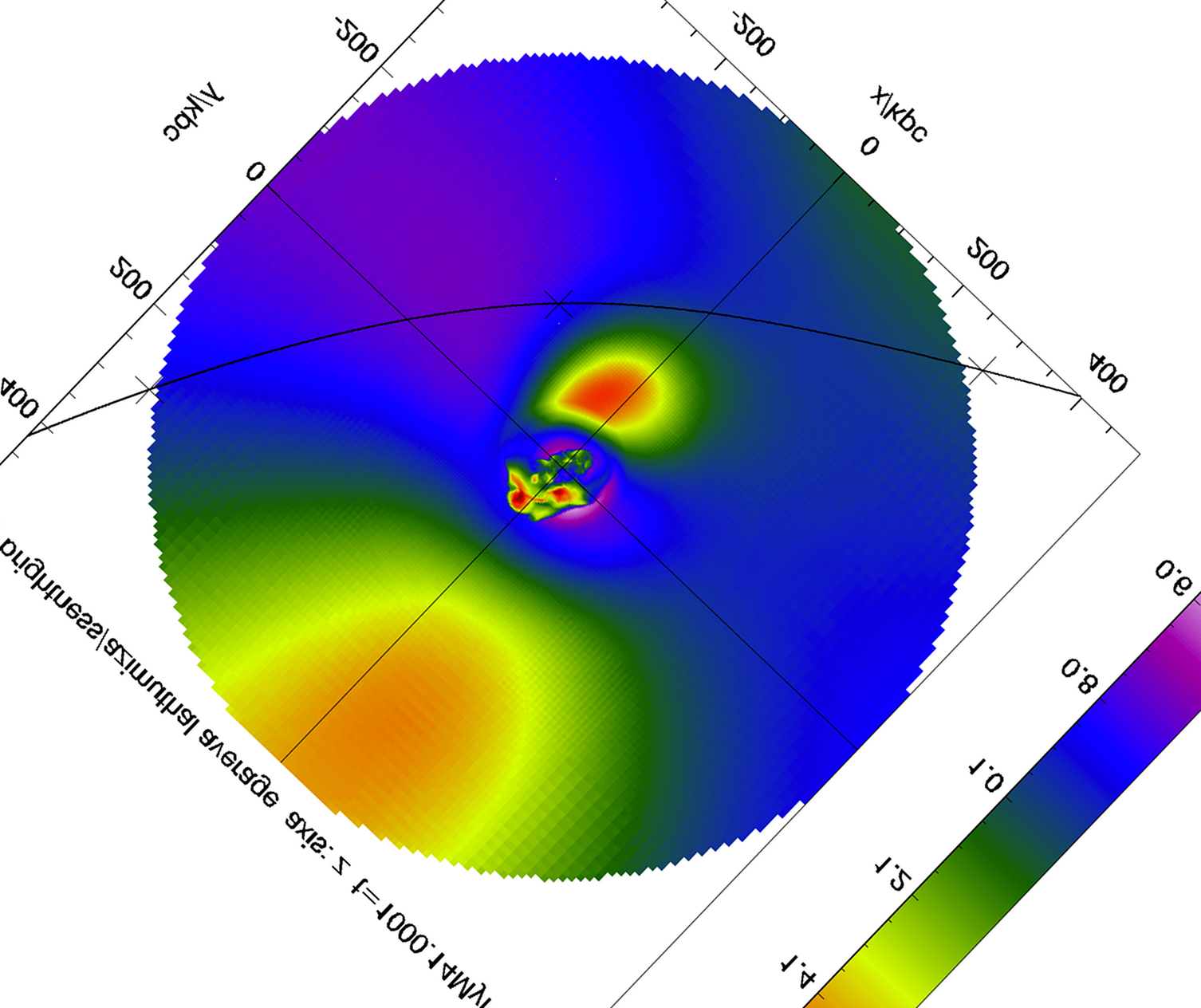}
\caption{Simulated SB residual map for A496 after 1 Gyr from the perycenter passage \citep{roediger12}, that we rotated to match the geometry of A2142.}
\label{fig:simul}
\end{figure}

The residual surface brightness images of A2142 shown in Fig.\,\ref{fig:residuals} can be compared with the predictions of numerical simulations to test the possibility that the fronts in A2142 are caused by sloshing. For instance, \citet{roediger11,roediger12} show predicted residual images from ad-hoc simulations of the Virgo cluster and A496 to reproduce the observed features and to infer the characteristics of the minor mergers that induced the sloshing. 
To perform a similar analysis on A2142, new tailored simulations are needed. 
A full hydro-dynamical N-body treatment will be required, because the rigid potential approximation \citep{roediger_zuhone} may not be completely valid in the outskirts (at 1 Mpc from the center, where we observe the SE cold front) . If the ellipticity of A2142 reflects the ellipticity of the potential well, the position and shapes of cold fronts may be different from spherical simulations.
Therefore, we limit the comparison with simulations at a qualitative level, and we refer for a more complete discussion to a forthcoming paper (Roediger et al. in prep.). \\
In Fig.~\ref{fig:simul}, we show a simulated residual image for A496 with the orbit of the perturber in the plane of the sky that we rotated to match the geometry of A2142. 
The morphological similarity between the observed and simulated maps is striking: they both show concentric excesses corresponding to the central cold fronts and a third excess at larger scale.
 In this geometry, the perturber should be moving in the west-east direction and should be now located in the east, but we do not have indication of a subcluster consistent with this orbit.
 Another possible match between observations and simulations could be obtained by rotating the residual map in Fig.~\ref{fig:simul} 180\deg, so that the simulated NW cold front would match the observed SE one, and the more central discontinuities would be the central cold fronts. However, we would expect a larger scale excess in the NW direction with this geometry, which is not seen in the residual maps obtained with either \xmm\ (Fig.~\ref{fig:residuals}) or ROSAT.\\
The main difference between observed and simulated maps is the larger scale of the sloshing phenomenon in A2142: cold fronts are located at about twice the distance from the center predicted by simulations. Cold fronts can be reproduced in simulations of massive clusters at those large scales if the sloshing phenomenon persists for about 2 Gyr or if they move outwards with a higher velocity, following a merger that is not very minor in terms of impact parameter, mass, and infall velocity of the subcluster). Another key difference is the presence of a sharp front delimiting the SE excess in A2142, while large scale features in simulations never show sharp discontinuities \citep{roediger11,roediger12}. As already mentioned,  full hydrodynamical N--body simulations are needed to characterize sloshing features at Mpc scales. However, we noticed that the surface brightness contrast in simulated central cold fronts depends on the velocity of the perturber. Therefore, the presence of a sharp front in the large scale excess in A2142 may be another indication of a merging event that is not very minor. However, the more central cold fronts in A2142, especially the NW one (Sec.~\ref{res_stability}), look remarkably smooth and stable, while all cold fronts in simulated violent mergers appear disturbed by the onset of hydrodynamical instabilities. A2142 is an interesting case for simulations to test the mechanisms which can suppress transport processes in the ICM and prevent the onset of Kelvin-Helmoltz instabilities, such as magnetic fields \citep{zuhone13a} and viscosity \citep{roediger13}. These mechanisms should be efficient at very different scales (from the few kpc scale of the central fronts to the Mpc scale of the SE one) and in different environments (both in the central dense regions of the cluster up to the rarefied outskirts). Moreover, they should be able to keep the fronts stable also in the hypothesis that they are due to an intermediate and not minor merger.  \\
The metal abundance and temperature distributions in A2142 (Sec.~\ref{sec:metal}) also  agree with predictions of simulations \citep{roediger11}: the richer and cooler gas is associated to the regions featuring a surface brightness excess. The gas in the large scale excess associated to the SE cold front is cooler but not significantly metal-richer than other regions. If we assume the metal abundance to remain constant during the sloshing process, the lower metallicity in the SE excess may suggest that its gas did not originate in the more central regions but at larger scales. This is indeed expected: \citet{ascas06}  showed that during sloshing the ICM does not change its distance from the center by more than a factor of three (see also Appendix~\ref{sec:app2}). 

\subsection{Comparison with optical results}
\label{sec:opt}
Recently, \citet{owers11} presented results based on multiobject spectroscopy of A2142. They selected 956 spectroscopically confirmed cluster members and provided evidence of significant substructures in the galaxy distribution. They identified several substructures and suggested  the subcluster S1 as the most likely perturber, which could have induced the sloshing. It consists of a small association of galaxies (7-10) around the second BCG (180 kpc NW of the first BCG) with a mean peculiar velocity $v_{pec}=1760$ km s$^{-1}$ and a dispersion $\sigma _v=224$ km s$^{-1}$, as returned by the KMM test. \citet{owers11} inferred that its motion is mainly aligned with our line of sight and that this orbit is consistent with the geometry of the two concentric cold fronts discovered by \chandra. However, the large scale residual map (Fig.~\ref{fig:residuals}, right panel) shows a hint of a spiral structure and a pattern more similar to the simulated images when the orbit of the perturber is in the plane of the sky (Fig.~\ref{fig:simul}). Nonetheless, it should be pointed out that the exact shape of the features in the residual maps depend on the underlying symmetry, and therefore, sloshing excesses may look different in a elliptical cluster and in a spherical symmetric one. Therefore, a perturber motion along the line of sight,  as suggested by \citet{owers11}, may still be consistent with our data, and it may also be possible that we are observing the sloshing in an intermediate inclination.\\
\citet{owers11} do not provide indication on the total mass of the candidate perturber and on the mass ratio of the merger, which could have induced the sloshing.
The small number of objects currently associated to this substructure should imply it to have a small mass, and therefore, the mass ratio should be rather small. However, during the perycenter passage, a subcluster suffers strong tidal forces during the perycenter passage, which could significantly reduce its galaxy concentration (e.~g.~\citealt{roediger12}). Indeed, the presence of a giant elliptical galaxy, which is the second ranked in the cluster with only a small difference in magnitude compared to the BCG ($\Delta M=0.3$ in the $K$ band), suggest that it could be the remnant of a more massive merging system whose less bound members were lost in an earlier phase of interaction, as originally pointed out by \citet{owers11}. We used the relation between central galaxy luminosity and the mass of the host halo \citep{lin04} to gauge the mass ratio of the merger.  Using the data from the Two Micron All-Sky-survey 
(2MASS \citealt{jarrett00, skrutskie06}) and following \citet{lin04}, we calculated the 
optical luminosities in the $K_S$ band for the two BCGs which are 
$6.18 \times 10^{11} L_{\odot}$ for 2MASX $J15582002+2714000$ and $4.71 \times 10^{11} L_{\odot}$ for 2MASX $J15581330+2714531$, assuming the luminosity 
distance for our adopted cosmology and redshift.
Using the \citet{lin04} relation for the whole BCG sample,
we can estimate the mass ratio by comparing the ratio of luminosities of the 
two BCGs: we expect a mass ratio of their parent halos of $0.35 \pm 0.06$ with the error bars just reflecting the statistical error on the slope of the relations\footnote{Instad, if we use the \citet{lin04} relation for the 82 clusters more massive  than
$10^{14} M_{\odot}$ we obtain a mass ratio $0.44 \pm 0.07$.}.  
These estimates are consistent
with an intermediate or major merger with a mass ratio on the order 
$1:3$ or even $1:2$. However, they should be considered 
only as a rough estimate, given the large scatter in the 
 $M_{halo}-L_{BCG}$ relation of \citet{lin04}, and we therefore cannot exclude lower
mass ratios typically used in sloshing simulations (e.~g.~\citealt{ascas06}).

\subsection{Comparison with radio results}
\label{sec:radio}

\begin{figure}
\includegraphics[width=\hsize]{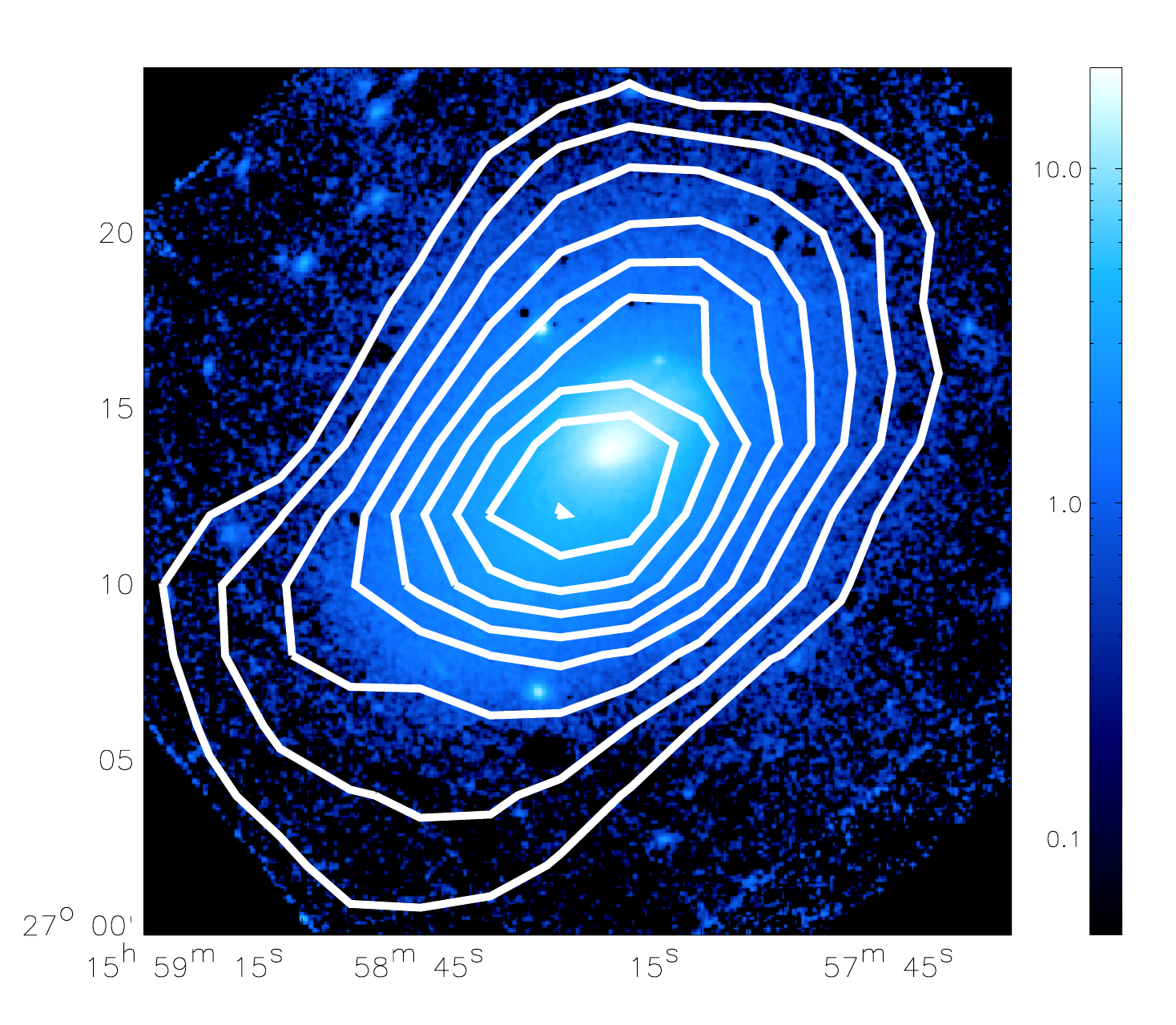}
\caption{GBT radio contours at $3,5,7$\ldots $\sigma$ level (Farnsworth et
al.\, 2013, in preparation) overlaid on the EPIC image.}
\label{fig:radio}
\end{figure}

The radio properties of A2142 make the overall picture even more intriguing.
\citet{gio00} detected extended $1.4$ GHz emission in the central region, which they classified as a mini--halo because of its small size. 
However, more recent $1.4$ GHz observations carried out with 
the Green Bank Telescope (GBT) show diffuse emission extending to the distant SE cold 
front and covering a size of $\sim$ 2 Mpc (Fig.\,\ref{fig:radio}). We refer to \citet{farnsworth13} for details on the observation and data reduction. This giant
radio halo has very low surface brightness with the peak in the GBT image
being $\sim 0.2\mu$Jy arcsec$^{-2}$ (45 mJy/b), and it is not surprising it has 
escaped from previous observations. In Fig.~\ref{fig:radio}, we show that the radio contours are elongated in the SE-NW direction similar to X-rays, but more detailed morphological comparison are hampered by the poor resolution of GBT data. \\% and suggest the possibility that the radio emission may be associated with the SE cold front. Indeed, 
Comparing the dynamical state of A2142 (Sec. \ref{sec:maps_morph}) with the statistical properties of the sample of clusters presented in \citet{cassano10}, we note that A2142 is more relaxed than the typical clusters hosting giant radio halos with the possible exception of A697 \citep{macario10}. In both A2142 and A697, we have indications for a merger mainly along the line of sight, which could trigger the radio emission without significantly disturbing the X-ray morphology.
Central sloshing in relaxed cool core clusters has been invoked as a possible re--acceleration mechanism for the formation of
mini--halo cluster sources \citep{mazzotta08,zuhone12}. It is tempting to consider that the same re-acceleration mechanisms may be also efficient outside the dense cores of galaxy clusters and that large scale sloshing (and not only major mergers) may trigger Mpc-scale radio emission. Sloshing could create a sort of ``giant mini-radio halo'', without dramatically disturbing the X-ray morphology. Alternately, the radio-halo may have been generated by the same intermediate merging event which induced the sloshing. In this case, we would expect the spectrum to be very steep, since the less energetic merger would have induced a lower level of turbulence in the ICM, resulting in less efficient particle acceleration phenomena \citep{brunetti08}. New observations are needed to distinguish between these scenarios. 
Indeed, GBT data do not allow to verify a possible association of the radio emission with the cold fronts in this cluster, as observed in mini--halos. Higher resolution VLA ($1.4$ GHz) and GMRT observations (240, 325, and 610 MHz) will soon allow us to verify this intriguing possibility.  Observations at lower frequencies could also provide the necessary information to measure the spectral index of this halo and to verify if it is consistent with typical halos and mini--halos or if the low brightness emission detected by GBT is just the faint end of a new Ultra Steep Spectrum Radio halo \citep{brunetti08}.\\

\subsection{Comparison with other sloshing cold fronts}
\label{sec:compare}
\begin{figure}
\includegraphics[width=\hsize]{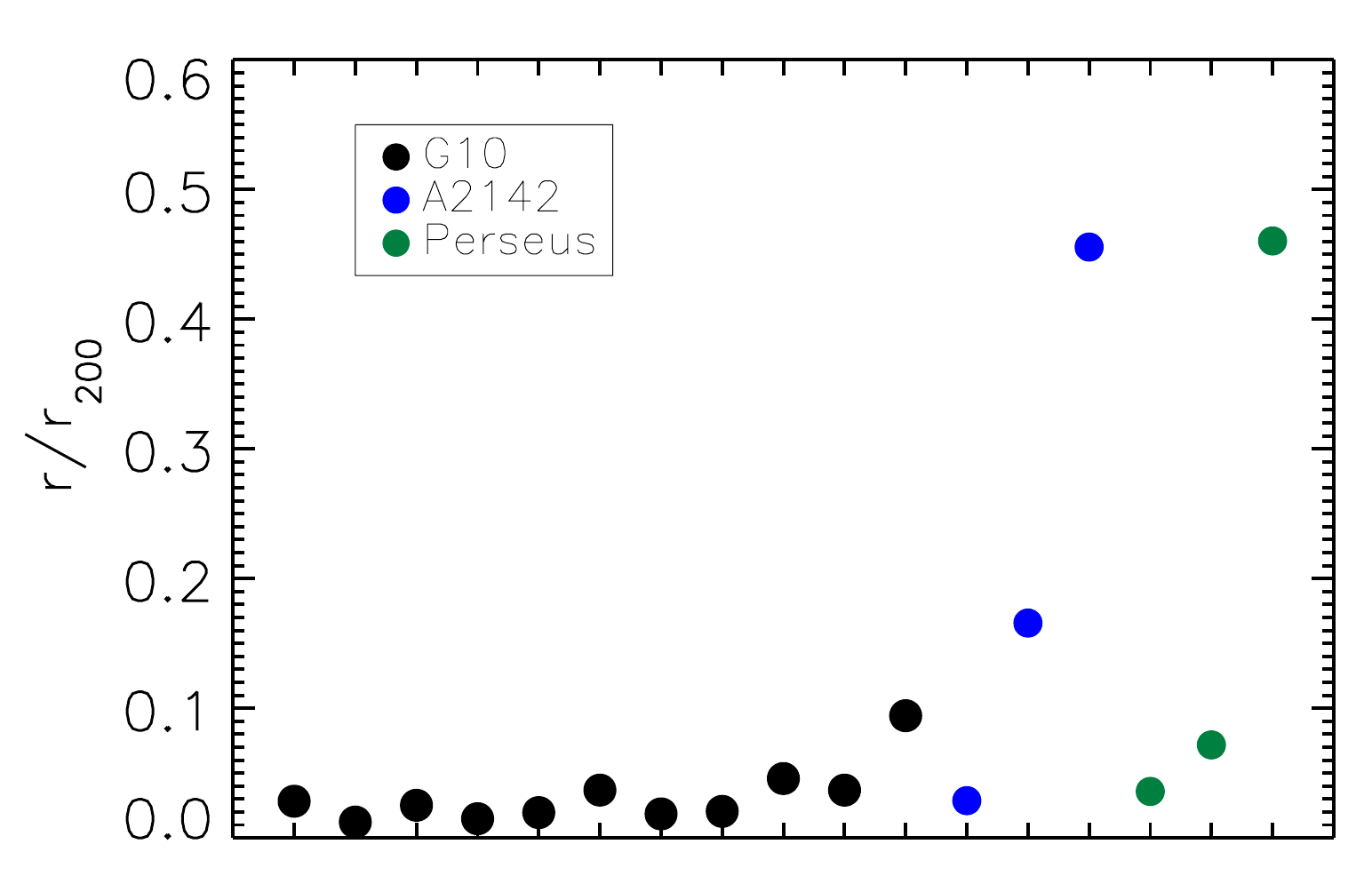}
\caption{Position of cold fronts in A2142 compared with those of sloshing cold fronts in the sample of \citealt{ghizzardi10} (G10) and in Perseus.}
\label{fig:poscf}
\end{figure}

Sloshing cold fronts have been observed so far mainly in the more central regions of cool core clusters within $0.1r_{200}$, where the entropy profile is steep, within (\citealt{ghizzardi10}, see also Fig.\,\ref{fig:poscf}). Surface brightness excesses at larger scales (up to 500 kpc), predicted by numerical simulations \citep[e.~g.~][]{roediger12}, are now observed in a few clusters (e.~g.~ A496, \citealt{roediger12} and references therein, and A2052, \citealt{blanton11}).
Recently, \citet{simio12} analyzed residual images obtained with the \rosat/PSPC and \xmm\ mosaic of the Perseus cluster and found alternating surface brightness excesses in the East-West  direction. The eastern excess is delimited at 700 kpc from the center by a cold front, detected as a surface brightness and temperature discontinuity in Suzaku profiles, while the western excess extends beyond 1 Mpc from the center but does not end with a cold front. The shape and position of the new features discovered in Perseus by \citet{simio12} are similar to what we have found in A2142: in both cases, the presence of multiple and concentric excesses suggest that the ICM of the cluster is sloshing at scales larger than observed so far and than predicted by simulations.\\
In Fig.\,\ref{fig:poscf}, we compare the positions of the three edges in A2142 with those of bona-fide sloshing cold fronts in \citet{ghizzardi10} and those in Perseus \citep{simio12}. We compare them in units of $r_{200}$ to account for the dependance on the total mass of the cluster. The external Perseus and A2142 cold fronts (but also the NW one in  A2142) are located at a larger distance in terms of fraction of the virial radius than the classical sloshing cold fronts in the \citet{ghizzardi10} sample. The range $(0.1-0.25)R_{200}$ was indeed explored in the sample of \citealt{ghizzardi10} (at a median redshift of $0.04$, the \xmm\ FOV covers about 30\% of $R_{200}$ for a $5$ keV cluster), but cold fronts found in this range all belong to the ``remnant-core'' class. \\
To our knowledge, the southeastern cold front in A2142 and the one in Perseus are the only examples, where the sloshing phenomenon is not confined in the central regions.  The phenomenon extends well beyond the cooling radius to $0.5r_{200}$ and involves a large fraction of the ICM.
A systematic research of these large scale features, similar to the one performed at smaller scales by \citet{ghizzardi10}, is necessary to assess their occurrence in the population of galaxy clusters and to establish if they are a common feature of the sloshing phenomenon or if the clusters that host them are rare and peculiar objects (Rossetti et al. in prep.).  

\subsection{Energy associated to the large-scale sloshing}
\label{sec:energy}
We developed a method to estimate the work associated to the sloshing against the gravitational potential to move a gas bubble  from a radius $r_o$ to a radius $r$ in a spherically symmetric potential well. We know from simulations (e.~g.~\citealt{ascas06}) that gas is not displaced all the way from the center out to the cold fronts, but it undergoes only smaller scale motions. To quantify these motions we used a basic idea: assuming that all transformations are adiabatic, we can compare the entropy values in the region of the cold fronts with the mean entropy profile of the cluster to reconstruct from where the gas originates. The work associated to the sloshing can be expressed as
\begin{equation}
E_{tot}=\sum_{r_i\in \mbox{\scriptsize{excess}}} M_{gas}(r_i)\int_{c(r_i)}^{r_i} g(r) \left(1-\left[\frac{K(c(r))}{K(r)}\right]^{3/5}\right)\, dr,
%\label{eq:final}
\end{equation}
where the summation is over all radii $r_i$ that correspond to excess emission in the residual map; $M_{gas}(r_i)$ is the gas mass in the $i$-th bin of the sloshing gas mass profile (Appendix \ref{sec:app2}); $c(r_i)$ is the starting radius for each gas particle at radius $r_i$; $g(r)$ is the modulus of the gravity acceleration; and $K(r)$ is the entropy at radius $r$. Details of the derivation of Eq.~1 are provided in Appendix \ref{sec:app2}. With this equation, we find $E_{tot}\sim3.5\times10^{61}$ ergs. We also estimated a lower limit to the sloshing work (see also Appendix \ref{sec:app2}) to account for the effect of the large ellipticity, and we found $ E_{tot} \sim 1.1\times10^{61} \mbox{ ergs}$. \\ 
We underline that the energy we estimated in this way is a fraction of the total energy associated to the sloshing. We can compare it with a rough estimate of the kinetic energy of the sloshing gas: we assume that all gas in the excess region, with mass $M_{gas}=\sum_iM_{gas}(r_i)$, is moving subsonically with a Mach number in the range of $0.3-0.5$ and we find that $E_{kin}=3.4-9\times10^{61}$ ergs. Unfortunately, we do not have a more precise measurement of gas velocity in sloshing, and the value we found is likely overestimated because not all the gas in the excess region is actually moving at these velocities. This result shows that the kinetic energy is comparable or slightly larger (a factor of 3-4) than  the energy we found in Eq.~1. In any case, we can treat the work calculated with Eq.~1 as a lower limit to the energy associated to the sloshing, since it is the minimum amount of energy needed to displace the gas particles at their current location. \\
We can compare the minimum energy associated to the sloshing with the total thermal energy of the ICM within the same radial range, $E_{th}(<1\,\rm{Mpc})\simeq 1.4 \times 10^{63}$ ergs . Therefore, the  minimum energy associated to the motion of the sloshing ICM is only a low fraction ($1-3\%$) of the thermal energy of the gas, even when gas motions involve a significant fraction of the ICM up to $0.45r_{200}$ as in the case of A2142. This fraction increases to almost 10\%, if we also consider the kinetic energy, which is consistent with the prediction of numerical simulations (Figure 24 in \citealt{zuhone12}).\\
The result discussed above implies that the minimum input energy necessary for the onset of the sloshing mechanism, even when it reaches very large scales, is not huge. Compared to the total energy of major mergers ($10^{63}-10^{64}$ ergs, e.~g.~\citealt{sarazin02} ) our estimate is almost negligible. However, the fraction of the merger energy that is spent in the different physical processes associated to the interaction is not completely assessed yet. 
If the above fraction is on the order of a few to ten per cent, the value we recovered is consistent with the minor merger scenario usually associated to the sloshing, but it may also be compatible with a more violent event. 

\subsection{A2142: an extreme case for sloshing?}
\label{sec:extreme}
The large scale features observed in both A2142 and Perseus cannot be completely explained within the current scenario of the sloshing phenomenon.
On the observational side, A2142 is a very different case from the other clusters undergoing sloshing: it not only hosts a cold front at almost 1 Mpc from the center, but also both its morphology and thermodynamic quantities suggest that it is not a completely relaxed cool-core system. Moreover, it hosts a giant radio halo. 
It is tempting to consider that these characteristics may be connected to each other and that the cause of the large scale sloshing (Sec.~\ref{sec:simul}) may have had other effects in the cluster, such as destroying the cool core and accelerating particles on $\simeq$ Mpc scales. \\
This tempting connection between the peculiar characteristics of A2142, may have wider implications for the physics of the ICM. Indeed, the origin of the CC-NCC dichotomy is still debated in the literature (see \citealt{rossetti10} and references therein), and there are a few clusters where the absence of a cool core is not clearly associated to indications of a major merger. The sloshing mechanism has been suggested as a possible mechanism to quench the cooling in cool core clusters, and numerical simulations have shown that it could be  sufficient, provided that the merging rate is high enough \citep{zuhone10}. The effect of large scale sloshing on the thermodynamic quantities of the ICM needs to be verified with dedicated simulations to test the possibility that sloshing may increase the entropy of the core to the level observed in A2142. \\
As discussed in Sec.~\ref{sec:radio}, sloshing may also prove an intriguing mechanism to accelerate particles to radio-emitting energies on scales larger than the central dense regions where mini radio halos are usually observed. Higher resolution radio observations are necessary to verify the presence of a spatial correlation between the radio emission and the cold fronts and therefore, are crucial to test this intriguing hypothesis. \\
As discussed in Sec.~\ref{sec:simul}, tailored simulations are needed to assess the origin of the peculiar characteristics of the sloshing phenomenon in A2142.  The first possibility is that large scale cold fronts, such as those observed in A2142 and Perseus, may be the natural long term evolution of the gentle perturbation observed in many cool core objects. To explain the presence of the giant radio halo and the absence of a cool core in this scenario, we have to assume that sloshing may have had ``extreme'' effects in the ICM of A2142, on both the thermal and  non-thermal population. Alternatively, we can interpret the large scale sloshing, the disruption of a cool core, the radio emission, and the presence of two BCGs with similar luminosity as indirect evidence of an ``intermediate'' merger. This merger may not be a major one but should be more violent than the typical minor off-axis interactions usually associated to sloshing. Present data and the existing simulations do not allow to clearly distinguish between these two scenarios, but both interpretations of A2142 can indeed be considered an ``extreme case'' for sloshing. 

\section{Summary and Conclusions}
We presented results obtained with a \xmm\ observation of A2142. This cluster could be considered the archetype object with multiple cold fronts, but its properties at large scales (beyond the \chandra\ field of view) are presented here for the first time, using a proprietary \xmm\ and archival \rosat observations. 
We discovered a surface brightness discontinuity at almost 1 Mpc SE from the cluster center. The spectral analysis inside and outside the edge has allowed us to exclude the possibility that this newly detected feature is a shock at more than 5$\sigma$ and to confirm that it is another cold front, the fourth in A2142 and the most distant one ever detected in a galaxy cluster. The overall geometry of the multiple edges, the residual surface brightness image, and the thermodynamic and metal abundance distribution qualitatively agree with predictions of numerical simulations of the sloshing phenomenon \citep{roediger11,roediger12}. However, the sloshing here occurs at much larger scales compared to simulations and to most clusters with sloshing cold fronts. The notable exception is Perseus, which also exhibits large scale ($>700$ kpc) features \citep{simio12}. At the same time, A2142 is different from typical sloshing clusters: it cannot be classified as a cool-core, and it hosts a newly discovered giant radio halo, although its morphology is regular, which makes it an unique system. These extreme characteristics of A2142 may be indirect effects of an intermediate merger, more violent than the typical minor off-axis interactions which induce gentle perturbations in the cores of relaxed clusters, but less disruptive than major mergers. \\
The observations of cold fronts beyond the more central regions in A2142 and in Perseus, and the large scale excesses observed in other clusters (e.g. A496, \citealt{roediger12}, A2052, \citealt{blanton11}) revolutionize our understanding of the sloshing phenomenon.
Sloshing does not involve only the densest and coolest gas in the cores of relaxed clusters, but it is a cluster-wide phenomenon, which can move high fractions of the ICM out to $R_{500}$. As such, it may have strong effects on the global properties of the cluster, such as destroying the cool core signature and inducing cluster-scale radio emission. \\
Numerical simulations so far have focused mainly on reproducing cold fronts and the effects of sloshing in the more central regions of galaxy clusters. New tailored simulations are certainly needed to reproduce large scale features, to understand under which conditions they develop, and to assess the effects that large scale sloshing may have on the thermal and non-thermal properties of the cluster. At the same time, it will be important to assess whether A2142 and Perseus are peculiar and extreme cases or whether large scale cold fronts are a common phenomenon and they have been unobserved so far, since the focus has been mainly on the more central regions (Rossetti et al.~in prep.). 

\begin{acknowledgements}
We thank the referee for useful suggestions; Eugene Churazov for pointing out an error in our estimate of the sloshing energy; and Damon Farnsworth and Larry Rudnick for providing the unpublished radio contours in Fig.~9 and for useful discussions. 
M.R.~ acknowledges support by the program ``Dote Ricercatori per lo sviluppo del capitale umano nel Sistema Universitario Lombardo'', FSE 2007-2013, Regione Lombardia.
F.G.~ acknowledges financial contributions by the Italian Space Agency
through ASI/INAF agreements I/023/05/0 and I/088/06/0 for the data
analysis and I/032/10/0 for
the XMM-Newton operations. E.R.~acknowledges support by the Priority Programme 1573 (“Physics of the Interstellar Medium”) of the DFG (German Research Foundation) and the supercomputing grants NIC 4368 and 5027 at John von Neumann Institute for Computing (NIC) on the supercomputer JUROPA at Jülich Supercomputing Centre (JSC).
The present paper is based on observations obtained with \xmm, an ESA science mission with
instruments and contributions directly funded by ESA Member States and
the USA (NASA).
\end{acknowledgements}

\bibliographystyle{aa}
\bibliography{refs}

\begin{thebibliography}{49}
\expandafter\ifx\csname natexlab\endcsname\relax\def\natexlab#1{#1}\fi

\bibitem[{Anders \& Grevesse(1989)}]{anders89}
Anders, E. \& Grevesse, N. 1989, \gca, 53, 197

\bibitem[{{Ascasibar} \& {Markevitch}(2006)}]{ascas06}
{Ascasibar}, Y. \& {Markevitch}, M. 2006, \apj, 650, 102

\bibitem[{{Blanton} {et~al.}(2011){Blanton}, {Randall}, {Clarke}, {Sarazin},
  {McNamara}, {Douglass}, \& {McDonald}}]{blanton11}
{Blanton}, E.~L., {Randall}, S.~W., {Clarke}, T.~E., {et~al.} 2011, \apj, 737,
  99

\bibitem[{{Brunetti} {et~al.}(2008){Brunetti}, {Giacintucci}, {Cassano},
  {Lane}, {Dallacasa}, {Venturi}, {Kassim}, {Setti}, {Cotton}, \&
  {Markevitch}}]{brunetti08}
{Brunetti}, G., {Giacintucci}, S., {Cassano}, R., {et~al.} 2008, \nat, 455, 944

\bibitem[{{Cassano} {et~al.}(2010){Cassano}, {Ettori}, {Giacintucci},
  {Brunetti}, {Markevitch}, {Venturi}, \& {Gitti}}]{cassano10}
{Cassano}, R., {Ettori}, S., {Giacintucci}, S., {et~al.} 2010, \apjl, 721, L82

\bibitem[{{Cavagnolo} {et~al.}(2009){Cavagnolo}, {Donahue}, {Voit}, \&
  {Sun}}]{cava09}
{Cavagnolo}, K.~W., {Donahue}, M., {Voit}, G.~M., \& {Sun}, M. 2009, \apjs,
  182, 12

\bibitem[{{De Luca} \& {Molendi}(2004)}]{deluca03}
{De Luca}, A. \& {Molendi}, S. 2004, \aap, 419, 837

\bibitem[{{Eckert} {et~al.}(2012){Eckert}, {Vazza}, {Ettori}, {Molendi},
  {Nagai}, {Lau}, {Roncarelli}, {Rossetti}, {Snowden}, \&
  {Gastaldello}}]{eckert12}
{Eckert}, D., {Vazza}, F., {Ettori}, S., {et~al.} 2012, \aap, 541, A57

\bibitem[{{Ettori} {et~al.}(2002){Ettori}, {De Grandi}, \&
  {Molendi}}]{ettori02}
{Ettori}, S., {De Grandi}, S., \& {Molendi}, S. 2002, \aap, 391, 841

\bibitem[{{Farnsworth} {et~al.}(2013){Farnsworth}, {Rudnick}, {Brown}, \&
  {Brunetti}}]{farnsworth13}
{Farnsworth}, D., {Rudnick}, L., {Brown}, S., \& {Brunetti}, G. 2013, submitted
  to \apj

\bibitem[{{Ghizzardi} {et~al.}(2010){Ghizzardi}, {Rossetti}, \&
  {Molendi}}]{ghizzardi10}
{Ghizzardi}, S., {Rossetti}, M., \& {Molendi}, S. 2010, \aap, 516, A32+

\bibitem[{{Giovannini} \& {Feretti}(2000)}]{gio00}
{Giovannini}, G. \& {Feretti}, L. 2000, New Astronomy, 5, 335

\bibitem[{{Hashimoto} {et~al.}(2007){Hashimoto}, {B{\"o}hringer}, {Henry},
  {Hasinger}, \& {Szokoly}}]{hashimoto07}
{Hashimoto}, Y., {B{\"o}hringer}, H., {Henry}, J.~P., {Hasinger}, G., \&
  {Szokoly}, G. 2007, \aap, 467, 485

\bibitem[{{Jarrett} {et~al.}(2000){Jarrett}, {Chester}, {Cutri}, {Schneider},
  {Skrutskie}, \& {Huchra}}]{jarrett00}
{Jarrett}, T.~H., {Chester}, T., {Cutri}, R., {et~al.} 2000, \aj, 119, 2498

\bibitem[{{Kalberla} {et~al.}(2005){Kalberla}, {Burton}, {Hartmann}, {Arnal},
  {Bajaja}, {Morras}, \& {P{\"o}ppel}}]{kalberla05}
{Kalberla}, P.~M.~W., {Burton}, W.~B., {Hartmann}, D., {et~al.} 2005, \aap,
  440, 775

\bibitem[{{Leccardi} \& {Molendi}(2008)}]{leccardi08a}
{Leccardi}, A. \& {Molendi}, S. 2008, \aap, 486, 359

\bibitem[{{Leccardi} {et~al.}(2010){Leccardi}, {Rossetti}, \&
  {Molendi}}]{leccardi10}
{Leccardi}, A., {Rossetti}, M., \& {Molendi}, S. 2010, \aap, 510, A82+

\bibitem[{{Lin} \& {Mohr}(2004)}]{lin04}
{Lin}, Y.-T. \& {Mohr}, J.~J. 2004, \apj, 617, 879

\bibitem[{{Macario} {et~al.}(2010){Macario}, {Venturi}, {Brunetti},
  {Dallacasa}, {Giacintucci}, {Cassano}, {Bardelli}, \& {Athreya}}]{macario10}
{Macario}, G., {Venturi}, T., {Brunetti}, G., {et~al.} 2010, \aap, 517, A43+

\bibitem[{{Markevitch} {et~al.}(2002){Markevitch}, {Gonzalez}, {David},
  {Vikhlinin}, {Murray}, {Forman}, {Jones}, \& {Tucker}}]{mark02b}
{Markevitch}, M., {Gonzalez}, A.~H., {David}, L., {et~al.} 2002, \apjl, 567,
  L27

\bibitem[{{Markevitch} {et~al.}(2000){Markevitch}, {Ponman}, {Nulsen}, {Bautz},
  {Burke}, {David}, {Davis}, {Donnelly}, {Forman}, {Jones}, {Kaastra},
  {Kellogg}, {Kim}, {Kolodziejczak}, {Mazzotta}, {Pagliaro}, {Patel}, {Van
  Speybroeck}, {Vikhlinin}, {Vrtilek}, {Wise}, \& {Zhao}}]{mark00}
{Markevitch}, M., {Ponman}, T.~J., {Nulsen}, P.~E.~J., {et~al.} 2000, \apj,
  541, 542

\bibitem[{{Markevitch} \& {Vikhlinin}(2007)}]{mark07}
{Markevitch}, M. \& {Vikhlinin}, A. 2007, \physrep, 443, 1

\bibitem[{{Markevitch} {et~al.}(2003){Markevitch}, {Vikhlinin}, \&
  {Forman}}]{mark03}
{Markevitch}, M., {Vikhlinin}, A., \& {Forman}, W.~R. 2003, in Astronomical
  Society of the Pacific Conference Series, Vol. 301, Astronomical Society of
  the Pacific Conference Series, ed. {S.~Bowyer \& C.-Y.~Hwang}, 37

\bibitem[{{Markevitch} {et~al.}(2001){Markevitch}, {Vikhlinin}, \&
  {Mazzotta}}]{mark01a}
{Markevitch}, M., {Vikhlinin}, A., \& {Mazzotta}, P. 2001, \apjl, 562, L153

\bibitem[{{Maughan} {et~al.}(2008){Maughan}, {Jones}, {Forman}, \& {Van
  Speybroeck}}]{maughan08}
{Maughan}, B.~J., {Jones}, C., {Forman}, W., \& {Van Speybroeck}, L. 2008,
  \apjs, 174, 117

\bibitem[{{Mazzotta} \& {Giacintucci}(2008)}]{mazzotta08}
{Mazzotta}, P. \& {Giacintucci}, S. 2008, \apjl, 675, L9

\bibitem[{{Mazzotta} {et~al.}(2001){Mazzotta}, {Markevitch}, {Vikhlinin},
  {Forman}, {David}, \& {VanSpeybroeck}}]{mazzotta01}
{Mazzotta}, P., {Markevitch}, M., {Vikhlinin}, A., {et~al.} 2001, \apj, 555,
  205

\bibitem[{{Oegerle} {et~al.}(1995){Oegerle}, {Hill}, \& {Fitchett}}]{oegerle95}
{Oegerle}, W.~R., {Hill}, J.~M., \& {Fitchett}, M.~J. 1995, \aj, 110, 32

\bibitem[{{Okabe} \& {Umetsu}(2008)}]{okabe08}
{Okabe}, N. \& {Umetsu}, K. 2008, \pasj, 60, 345

\bibitem[{{Owers} {et~al.}(2011){Owers}, {Nulsen}, \& {Couch}}]{owers11}
{Owers}, M.~S., {Nulsen}, P.~E.~J., \& {Couch}, W.~J. 2011, \apj, 741, 122

\bibitem[{{Owers} {et~al.}(2009){Owers}, {Nulsen}, {Couch}, \&
  {Markevitch}}]{owers09}
{Owers}, M.~S., {Nulsen}, P.~E.~J., {Couch}, W.~J., \& {Markevitch}, M. 2009,
  \apj, 704, 1349

\bibitem[{{Roediger} {et~al.}(2011){Roediger}, {Br{\"u}ggen}, {Simionescu},
  {B{\"o}hringer}, {Churazov}, \& {Forman}}]{roediger11}
{Roediger}, E., {Br{\"u}ggen}, M., {Simionescu}, A., {et~al.} 2011, \mnras,
  413, 2057

\bibitem[{{Roediger} {et~al.}(2013){Roediger}, {Kraft}, {Forman}, {Nulsen}, \&
  {Churazov}}]{roediger13}
{Roediger}, E., {Kraft}, R.~P., {Forman}, W.~R., {Nulsen}, P.~E.~J., \&
  {Churazov}, E. 2013, \apj, 764, 60

\bibitem[{{Roediger} {et~al.}(2012){Roediger}, {Lovisari}, {Dupke},
  {Ghizzardi}, {Br{\"u}ggen}, {Kraft}, \& {Machacek}}]{roediger12}
{Roediger}, E., {Lovisari}, L., {Dupke}, R., {et~al.} 2012, \mnras, 420, 3632

\bibitem[{{Roediger} \& {Zuhone}(2012)}]{roediger_zuhone}
{Roediger}, E. \& {Zuhone}, J.~A. 2012, \mnras, 419, 1338

\bibitem[{{Rossetti} {et~al.}(2011){Rossetti}, {Eckert}, {Cavalleri},
  {Molendi}, {Gastaldello}, \& {Ghizzardi}}]{rossetti11}
{Rossetti}, M., {Eckert}, D., {Cavalleri}, B.~M., {et~al.} 2011, \aap, 532,
  A123

\bibitem[{{Rossetti} {et~al.}(2007){Rossetti}, {Ghizzardi}, {Molendi}, \&
  {Finoguenov}}]{rossetti06}
{Rossetti}, M., {Ghizzardi}, S., {Molendi}, S., \& {Finoguenov}, A. 2007, \aap,
  463, 839

\bibitem[{{Rossetti} \& {Molendi}(2010)}]{rossetti10}
{Rossetti}, M. \& {Molendi}, S. 2010, \aap, 510, A83+

\bibitem[{{Santos} {et~al.}(2008){Santos}, {Rosati}, {Tozzi}, {B{\"o}hringer},
  {Ettori}, \& {Bignamini}}]{santos08}
{Santos}, J.~S., {Rosati}, P., {Tozzi}, P., {et~al.} 2008, \aap, 483, 35

\bibitem[{{Sarazin}(2002)}]{sarazin02}
{Sarazin}, C.~L. 2002, in ASSL Vol. 272: Merging Processes in Galaxy Clusters,
  1--38

\bibitem[{{Simionescu} {et~al.}(2012){Simionescu}, {Werner}, {Urban}, {Allen},
  {Fabian}, {Sanders}, {Mantz}, {Nulsen}, \& {Takei}}]{simio12}
{Simionescu}, A., {Werner}, N., {Urban}, O., {et~al.} 2012, \apj, 757, 182

\bibitem[{{Skrutskie} {et~al.}(2006){Skrutskie}, {Cutri}, {Stiening},
  {Weinberg}, {Schneider}, {Carpenter}, {Beichman}, {Capps}, {Chester},
  {Elias}, {Huchra}, {Liebert}, {Lonsdale}, {Monet}, {Price}, {Seitzer},
  {Jarrett}, {Kirkpatrick}, {Gizis}, {Howard}, {Evans}, {Fowler}, {Fullmer},
  {Hurt}, {Light}, {Kopan}, {Marsh}, {McCallon}, {Tam}, {Van Dyk}, \&
  {Wheelock}}]{skrutskie06}
{Skrutskie}, M.~F., {Cutri}, R.~M., {Stiening}, R., {et~al.} 2006, \aj, 131,
  1163

\bibitem[{{Snowden} {et~al.}(1998){Snowden}, {Egger}, {Finkbeiner}, {Freyberg},
  \& {Plucinsky}}]{snowden98}
{Snowden}, S.~L., {Egger}, R., {Finkbeiner}, D.~P., {Freyberg}, M.~J., \&
  {Plucinsky}, P.~P. 1998, \apj, 493, 715

\bibitem[{{Snowden} {et~al.}(1994){Snowden}, {McCammon}, {Burrows}, \&
  {Mendenhall}}]{snowden94}
{Snowden}, S.~L., {McCammon}, D., {Burrows}, D.~N., \& {Mendenhall}, J.~A.
  1994, \apj, 424, 714

\bibitem[{{Tittley} \& {Henriksen}(2005)}]{tittley05}
{Tittley}, E.~R. \& {Henriksen}, M. 2005, \apj, 618, 227

\bibitem[{{Vikhlinin} {et~al.}(2001){Vikhlinin}, {Markevitch}, \&
  {Murray}}]{vikh01b}
{Vikhlinin}, A., {Markevitch}, M., \& {Murray}, S.~S. 2001, \apj, 551, 160

\bibitem[{{ZuHone} {et~al.}(2013{\natexlab{a}}){ZuHone}, {Markevitch},
  {Brunetti}, \& {Giacintucci}}]{zuhone12}
{ZuHone}, J., {Markevitch}, M., {Brunetti}, G., \& {Giacintucci}, S.
  2013{\natexlab{a}}, \apj, 762, 78

\bibitem[{{ZuHone} {et~al.}(2010){ZuHone}, {Markevitch}, \&
  {Johnson}}]{zuhone10}
{ZuHone}, J.~A., {Markevitch}, M., \& {Johnson}, R.~E. 2010, \apj, 717, 908

\bibitem[{{ZuHone} {et~al.}(2013{\natexlab{b}}){ZuHone}, {Markevitch},
  {Ruszkowski}, \& {Lee}}]{zuhone13a}
{ZuHone}, J.~A., {Markevitch}, M., {Ruszkowski}, M., \& {Lee}, D.
  2013{\natexlab{b}}, \apj, 762, 69

\end{thebibliography}

\appendix
\section{Modeling of the surface brightness profile in the SE sector}
\label{app:sbprof}
\begin{figure*}
\begin{centering}
\begin{minipage}[t]{0.90\textwidth}
\resizebox{\hsize}{!} {
\includegraphics[angle=0,keepaspectratio,scale=1]{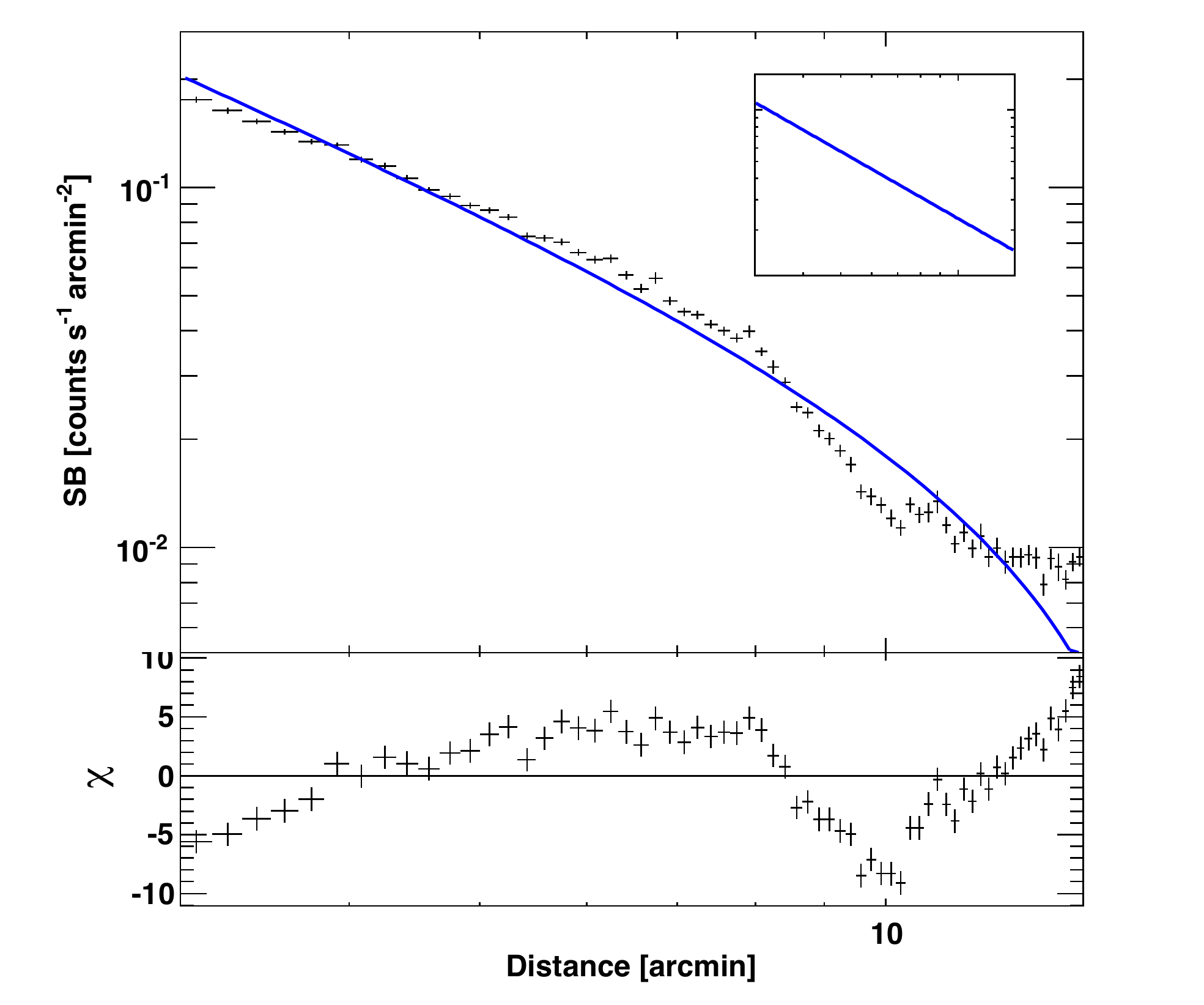}
%\caption{a}
\hspace{5mm}
\includegraphics[angle=0,keepaspectratio,scale=1]{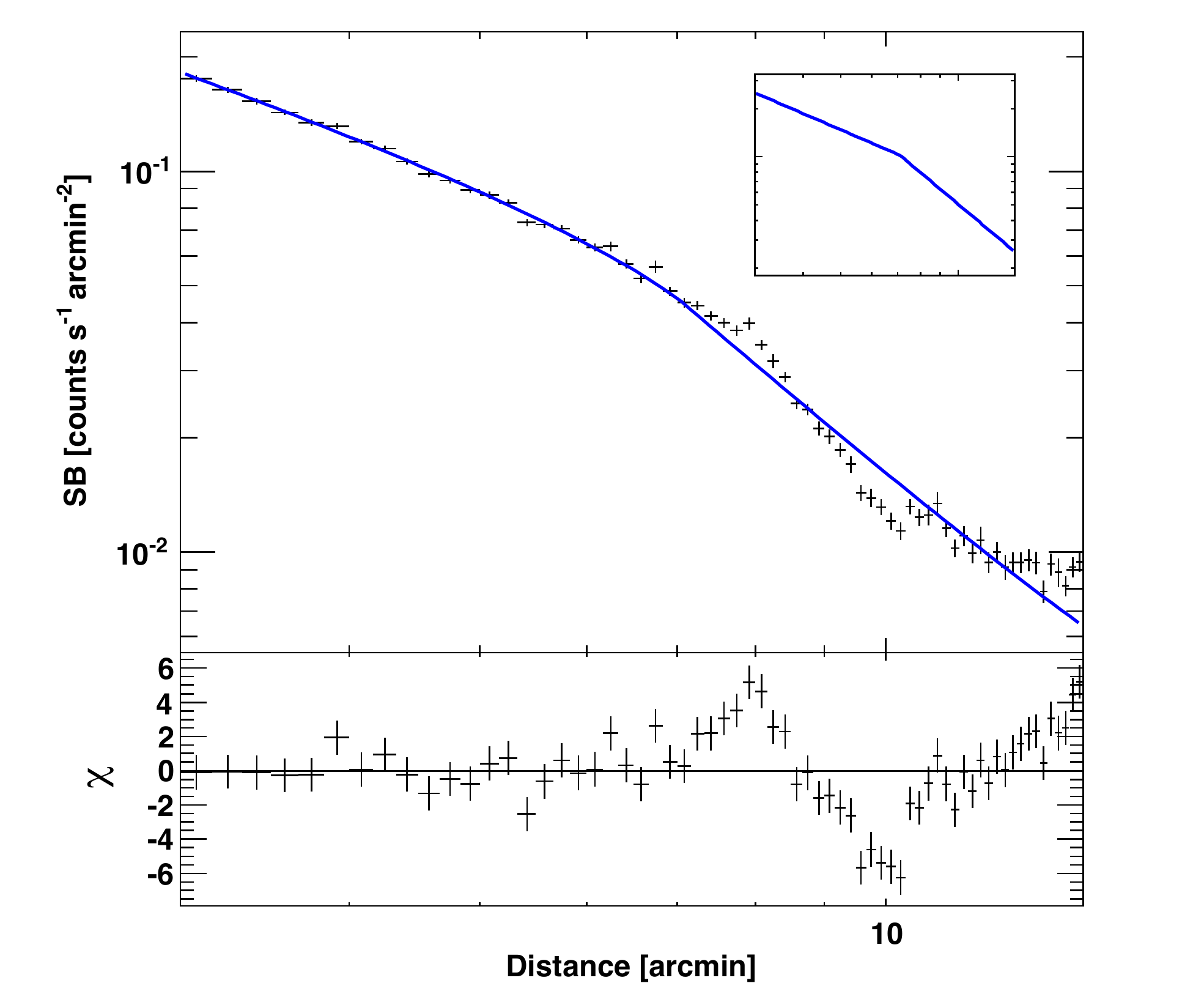}
}
\end{minipage}
\begin{minipage}[t]{0.90\textwidth}
\resizebox{\hsize}{!} {
\includegraphics[angle=0,keepaspectratio,scale=1]{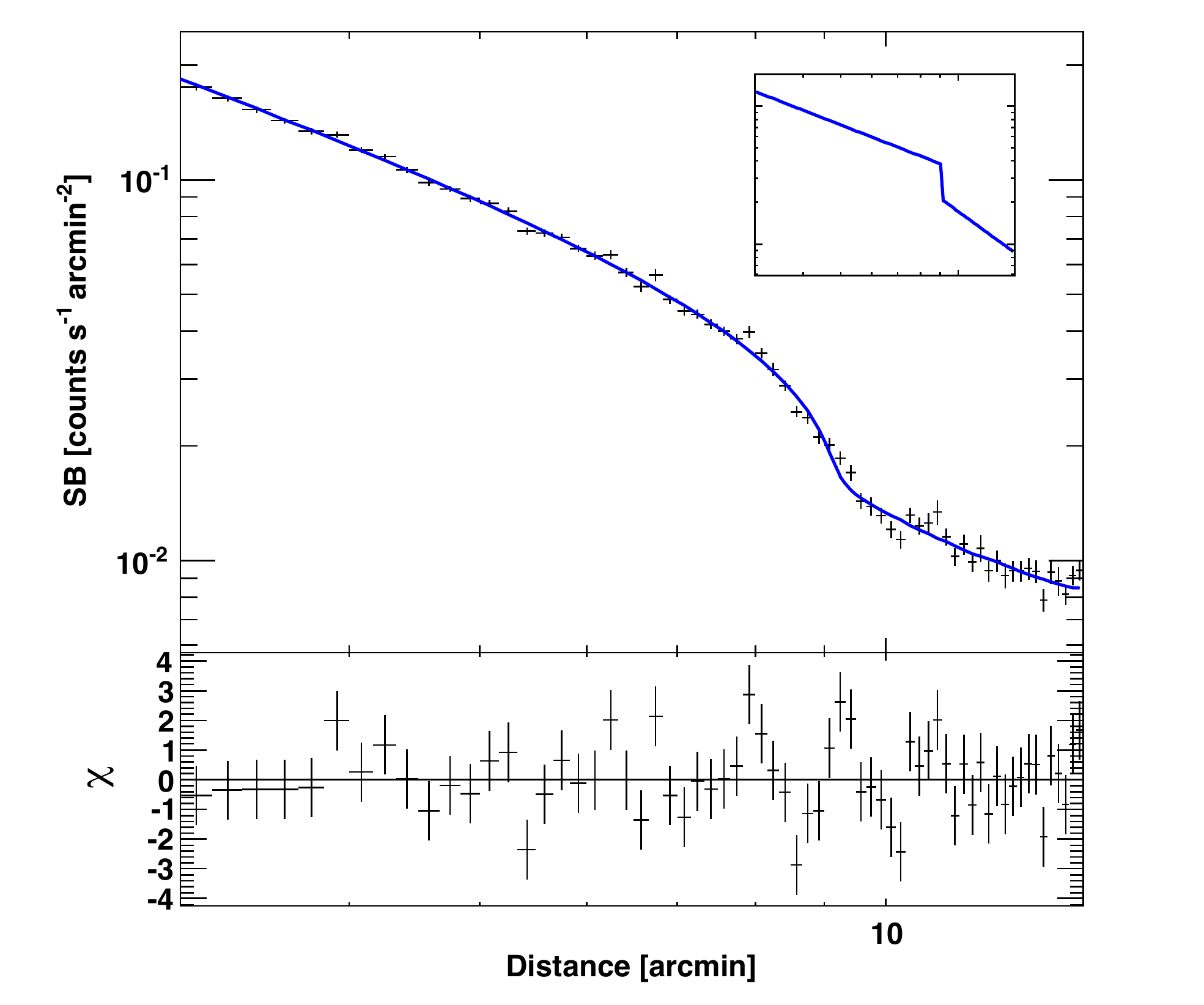}
%\caption{a}
\hspace{5mm}
\includegraphics[angle=0,keepaspectratio,scale=1]{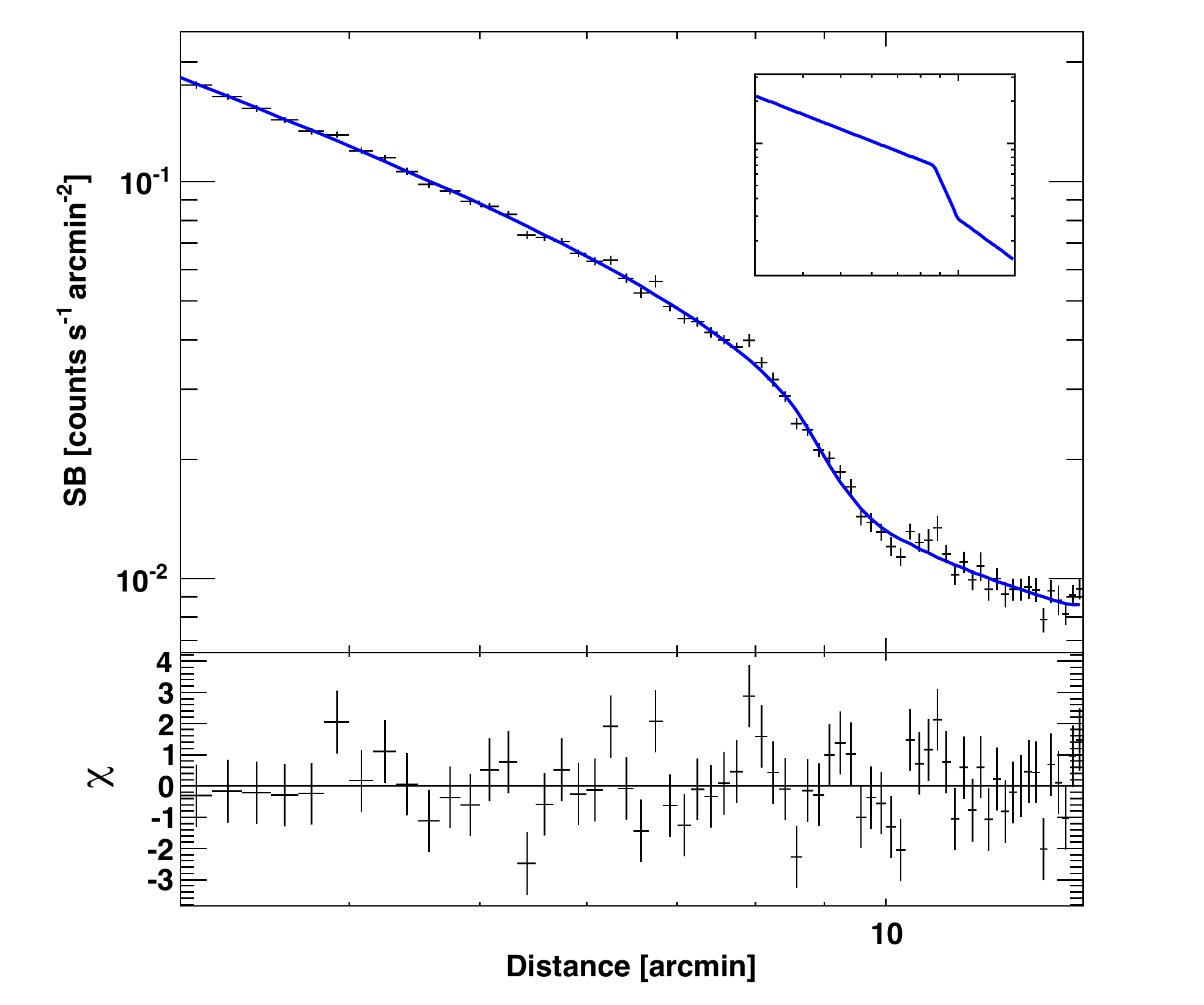}
}
\end{minipage}
\caption{EPIC/pn surface brightness profile in the SE sector (position angles 180-250$^\circ$) in the 0.5-2.0 keV band. The panel below each profile shows the residuals ($\Delta \chi^2$) of data from the best fit model. The insets show the density models used to fit the data: a single power-law ({\it upper left panel}), a double power-law  ({\it upper right panel}), a broken power-law ({\it lower left panel}), and a triple power-law  ({\it lower right panel}).}
\label{fig:app1}
\end{centering}
\end{figure*}

Since the SE large scale feature we detected in A2142 could be the first cold front observed with \xmm\ at a distance as large as 1 Mpc from the cluster center in low surface brightness regions, a detailed modeling of the profile is necessary to exclude other possible interpretations. We extracted the SB profile in the SE sector (position angles 180-250$^\circ$) from the EPIC/pn image in the $0.5-2$ keV energy range. We corrected for vignetting using the exposure map and we masked out bright point sources. We then assumed different models for the shape of the density profile that we project, assuming an elliptical symmetry as in \citet{owers09}, to fit the surface brightness data in the radial range $3-15 \arcmin$ after convolving with the \xmm\ PSF.
We started using two simple models that do not assume any discontinuity in the profile: a single power law, which would be the normal behavior of the density profile in the undisturbed outskirts of a cluster, and a double power law, which could model a steepening in the density profiles at  $r_{break}$. As shown in the upper panels in Fig.~\ref{fig:app1}, these simple models fail to reproduce the complex shape of the profile ($\chi^2_{red} = 17.79$ and $\chi^2_{red} = 6.26$ respectively). Then, we fitted the profile with the standard broken power-law density model used with cold fronts  (e.~g~ \citealt{owers09}):
\begin{equation}
n_e(r) = \left\{
  \begin{array}{l l}
    n_{1}\left(\frac{r}{r_{cut}}\right)^{-\alpha_{in}} & r < r_{cut} \\
    n_{2}\left(\frac{r}{r_{cut}}\right)^{-\alpha_{out}} & r \ge r_{cut}
  \end{array} \right. .
\label{eq:broken}
 \end{equation}
As discussed in Sec.~\ref{sec:se_ima}, the profile does not extend at large radii, so we fixed the slope of the outer component at the value $\alpha_{out}=2.05$ that we derived from the analysis of the ROSAT/PSPC profile.
This model reproduces the shape of the profile well (lower left panel in Fig.~\ref{fig:app1}). The best fit parameters are provided in Table \ref{tab:se_cf}.
 Although the fit with this model is satisfactory  ($\chi^2_{red} = 91.9/61\, \textrm{d.o.f.}=1.51$), we also used a fourth, more complicated density model, namely a combination of three power-laws:
\begin{equation}
n_e(r) = \left\{
  \begin{array}{l l}
    n_{1}r^{-\alpha_{1}} & r < r_{1} \\
    n_{2}r^{-\alpha_{2}} &  r_{1}< r < r_{2} \\
    n_{3}r^{-\alpha_{3}} &  r > r_{2}    
  \end{array} 
\right. ,
\label{eq:triple}
 \end{equation}
with the continuity condition at the breaks $r_1$ and $r_2$
\begin{eqnarray*}
n_1 \times r_1^{-\alpha_1}= n_2 \times r_1^{-\alpha_2}\\
n_2 \times r_2^{-\alpha_2}= n_3 \times r_2^{-\alpha_3}.
\end{eqnarray*}
We fixed $\alpha_3=2.05$, as in the previous case, and $r_{2}=10 \arcmin$ while we left as free parameters $\alpha_1$, $\alpha_2$, $r_1$ and the overall normalization. 
This model provides an improved fit of the data (lower right panel in Fig.~\ref{fig:app1}) with $\chi^2_{red} = 79.8/60\, \textrm{d.o.f.}=1.33$. It is interesting to note that the  model in Eq.~\ref{eq:broken} can be considered a limiting case for the triple power law model (Eq.~\ref{eq:triple}), where the slope $\alpha_2$ tends to infinity and the first break tends to coincide with the second one. In principle, we could compare the $\chi^2$ of the two fits and find that the triple power-law model provides a better description to the data than the double model at a confidence level of more than $99 \%$ ($\Delta \chi^2=12.1$) but this approach is limited by the fact that  it considers only statistical errors. Systematic uncertainties, such as those due to the assumption of an idealized geometry or to an imperfect modeling of the PSF, are likely dominant, and therefore, it is difficult to clearly distinguish between the two models. Moreover, the fit of the profile with Eq.~\ref{eq:triple},   requires the slope of the central power law to be extremely steep: $\alpha_2=6.2 \pm 0.5$ is higher both than the slopes in the inner and outer part of the profile ($\alpha_1=1.09 \pm 0.01$ and $\alpha_3=2.05$) and than the density slopes typically observed in the external regions of galaxy clusters ($\alpha$ in the range 2-3 \citealt{eckert12}). The fit also provides the position of the first break, $r_1=8.68 \pm 0.08$ arcmin, which implies that the region where the profile is described by the steep power-law is a thin shell of width 130 kpc. This best fit model describes a density distribution where the gas properties change very rapidly in a small shell, which is not much different from the sharp transition between two phases of the gas of the broken power law model (Eq.~\ref{eq:broken}), as clearly seen in the insets of the bottom panels in Fig.~\ref{fig:app1}. Therefore, the comparison of these two models supports the presence of a discontinuity in the SE profile of A2142.\\
The triple power-law model may also be explained by assuming that it is caused by a broken power-law model, where the discontinuity is not in the plane of the sky but has an inclination $\theta$. Projection of this model in this geometry induces a smoothing on the surface brightness profile, which in the plane of the sky would appear sharp. The smoothing scale can be used in combination of the azimuthal extension of the front to derive the inclination $\theta$ under cylindrical approximation (i.~e.~, assuming that the extension of the front along the line of sight is the same as in the plane of the sky). We used the width of the shell where the density profile is the described by the steep power law (130 kpc) as smoothing length and $1.2$ Mpc (Sec.~\ref{sec:se_ima}) as extension of the edge and we found an inclination of about $6^{\circ}$.
If the systematics were not dominating, we could therefore infer that the discontinuity is seen with a small, but non zero, angle with respect to the plane of the sky.

\section{Spectral background modeling in the MOS detectors}
\begin{figure}
\includegraphics[width=\hsize]{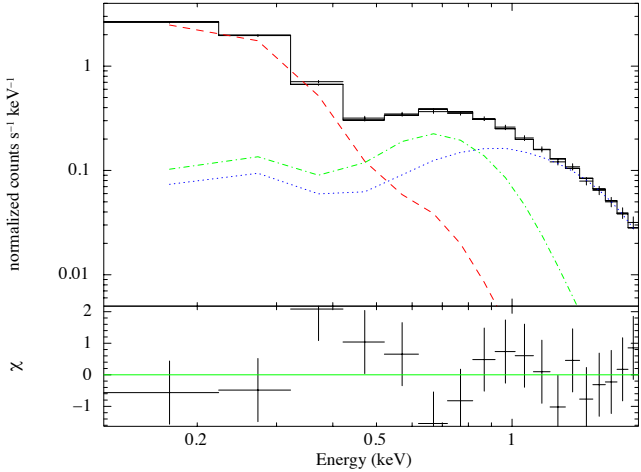}
\caption{\emph{ROSAT}/PSPC spectrum of the background region, that is located $50^\prime$ SE of the X-ray peak of A2142. The various components show the best-fit spectra for the local bubble (red), the Galactic halo (green), and the CXB (blue). }
\label{skybkg}
\end{figure}

 We applied a method similar to the one described in \citealt{leccardi08a} (hereafter LM08) to model the background in the two regions at the southeastern cold front and restricted our analysis to the MOS detectors, since their background was found to be more predictable than for the pn (see LM08). While LM08 used the outermost annulus of the MOS image to model the local sky background components (cosmic X-ray background or CXB, Galactic halo, and local hot bubble), the source in our case is filling the entire field of view, such that the sky components must be estimated from another instrument. Therefore, we used the \emph{ROSAT}/PSPC observation and extracted the spectrum of a background region located $50^\prime$ SE of the cluster core. We used the ftool \texttt{Xselect} to extract the spectrum of this region from the event files and computed the appropriate effective area using the \texttt{pcarf} tool. We subtracted the particle background with the help of the \texttt{pcparpha} executable. We fitted the resulting spectrum with a model composed by: i) a power law with a photon index fixed to 1.4 for the CXB \citep{deluca03}, absorbed by the Galactic $N_H$ \citep[$N_H=3.8\times10^{20}$ cm$^{2}$,][]{kalberla05}; ii) an absorbed APEC model with temperature fixed to 0.22 keV for the Galactic halo; and iii) an unabsorbed APEC model at a temperature of 0.11 keV for the local hot bubble \citep{snowden98}. The PSPC spectrum is well represented by this model (see Fig. \ref{skybkg}), and the normalization of the CXB component agrees with the value in \citet{deluca03}.\\ 
To model the non-X-ray background (NXB), we used the spectra from the unexposed areas of the corners of the MOS detectors and fitted them with a phenomenological model made of a broken power law and a number of Gaussian emission lines (see Appendix A of LM08). This function was then used to model the NXB spectrum for the IN and OUT regions (see Fig. \ref{fig:spec_inout}). The spectral shape of the NXB model was fixed to that found in the corners of the MOS, but the overall normalization of the model and the normalization of the prominent Al and Si emission lines were treated as free parameters. In both cases, the NXB dominates the source contribution above $\sim$ 6 keV; therefore, the normalization of the NXB component is very well constrained by the high-energy data. The sky components were taken from our \emph{ROSAT} background region (see Fig. \ref{skybkg}) and rescaled to the proper area, accounting for CCD gaps and dead pixels. The normalization of the local components was fixed, but the CXB component was allowed to vary by $\pm15\%$ to account for cosmic variance. In any case, we note that the sky components are relatively unimportant at all energies, so our results are unaffected by local variations of the sky background. In addition, we also modeled the spectral component from quiescent soft protons (QSPs), following the method described in Appendix B of LM08. Namely, we computed the ratio between the mean count rate inside and outside the FOV of the detectors and used the phenomenological formula derived by LM08 to derive the normalization of the QSP component. The spectral shape of the QSP component was fixed to that found in LM08.\\
Four offset observations in the outskirts of A2142 were recently perfomed by \xmm\ (P.~I. D.~Eckert) during AO11 (Eckert et al.~in preparation). We extracted spectra using these observations in external sectors to model the sky-background components and verify our estimate of the normalizations based on the \emph{ROSAT}/PSPC data. \xmm\ data should grant us a better spectral resolution and thus a more detailed spectral modeling. However, the spectra in the external regions are dominated by the instrumental background (much larger in \xmm\ than in \rosat) even at low energies. Nonetheless, we verified that the normalizations of the sky components with the \xmm\ offset observations are consistent with those we obtained with the \rosat/PSPC modeling and do not have a significant impact on the spectral parameters that we measured in Sec. \ref{sec:se_spe}. 

\section{Estimate of the sloshing energy}
\label{sec:app2}
\begin{figure*}
\resizebox{\hsize}{!}{\includegraphics[height=5cm]{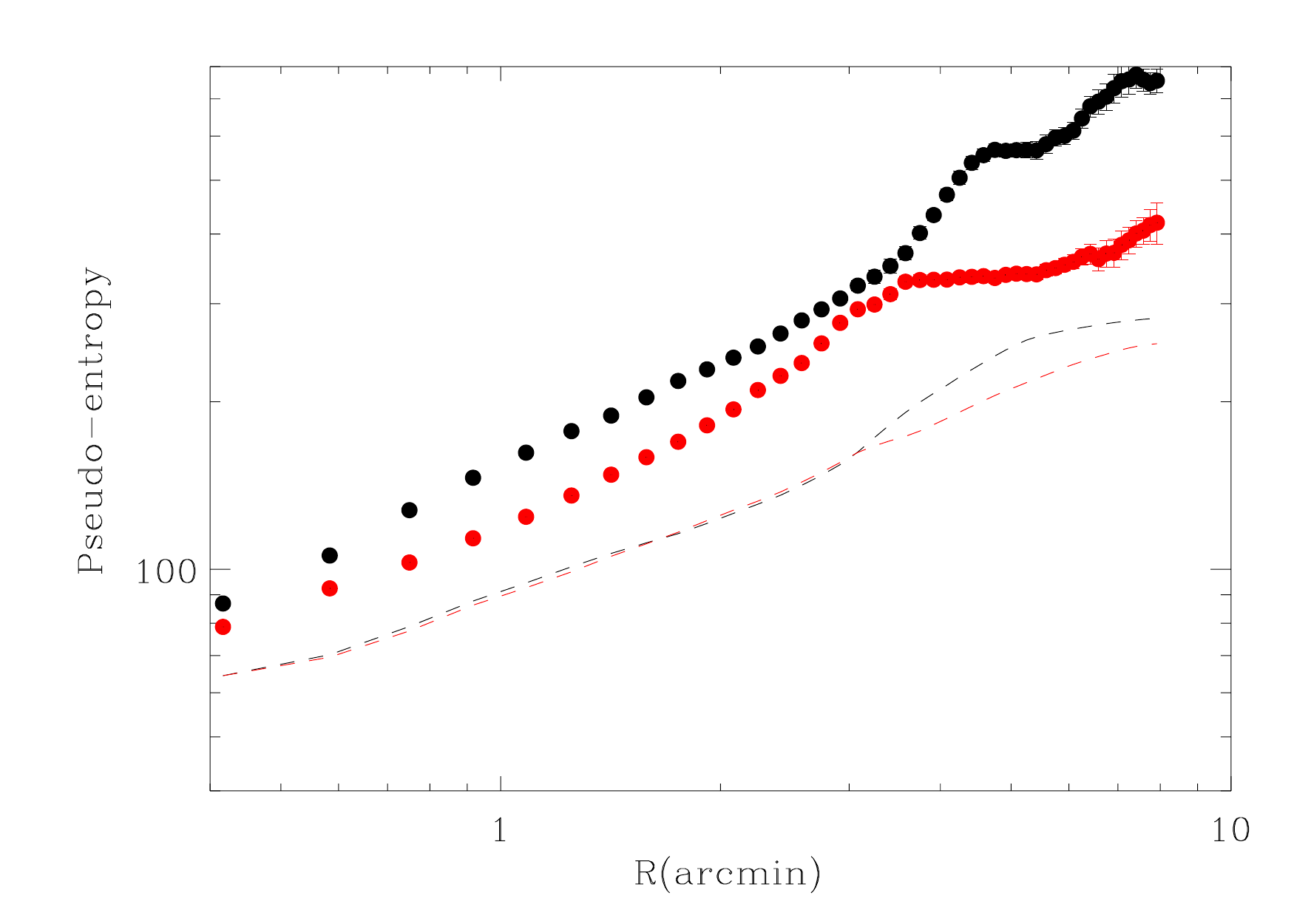}\includegraphics[height=5cm]{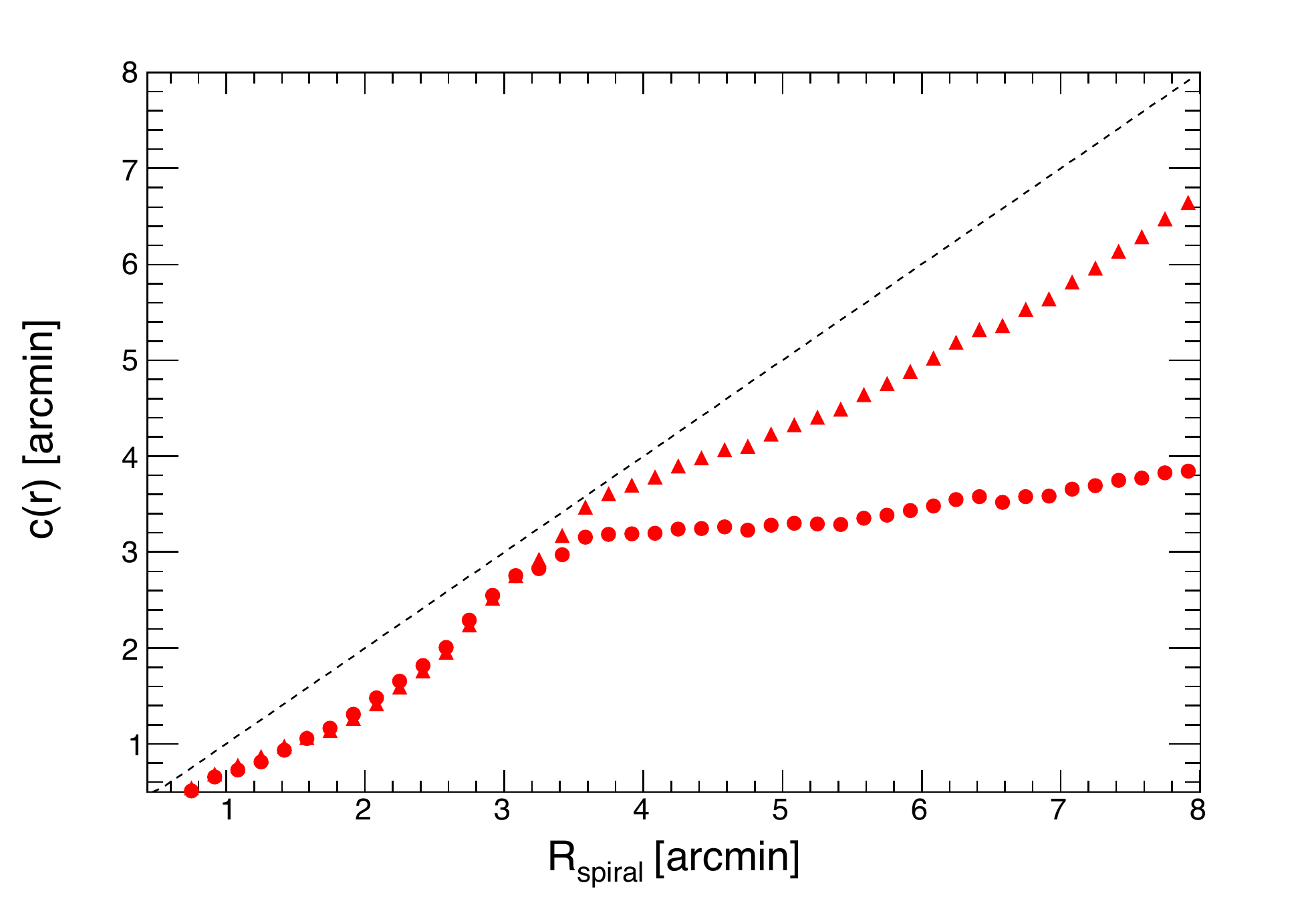}}
\caption{\emph{Left:} Pseudo-entropy profiles in the excess (red) and in the unperturbed region (black). Dashed lines show the entropy profiles in the elliptical simulation (rescaled for clarity) for the ``excess'' (red) and unperturbed (black) region. \emph{Right:} Corresponding radius $c(r)$ for which the pseudo-entropy in the unperturbed region is equal to that in the spiral region. Circles are obtained by neglecting the ellipticity, while triangles are the values corrected assuming that the ellipticity is entirely intrinsic. The dashed line shows the one-to-one relation.}
\label{entropy}
\end{figure*}

\begin{figure}
\includegraphics[width=\hsize]{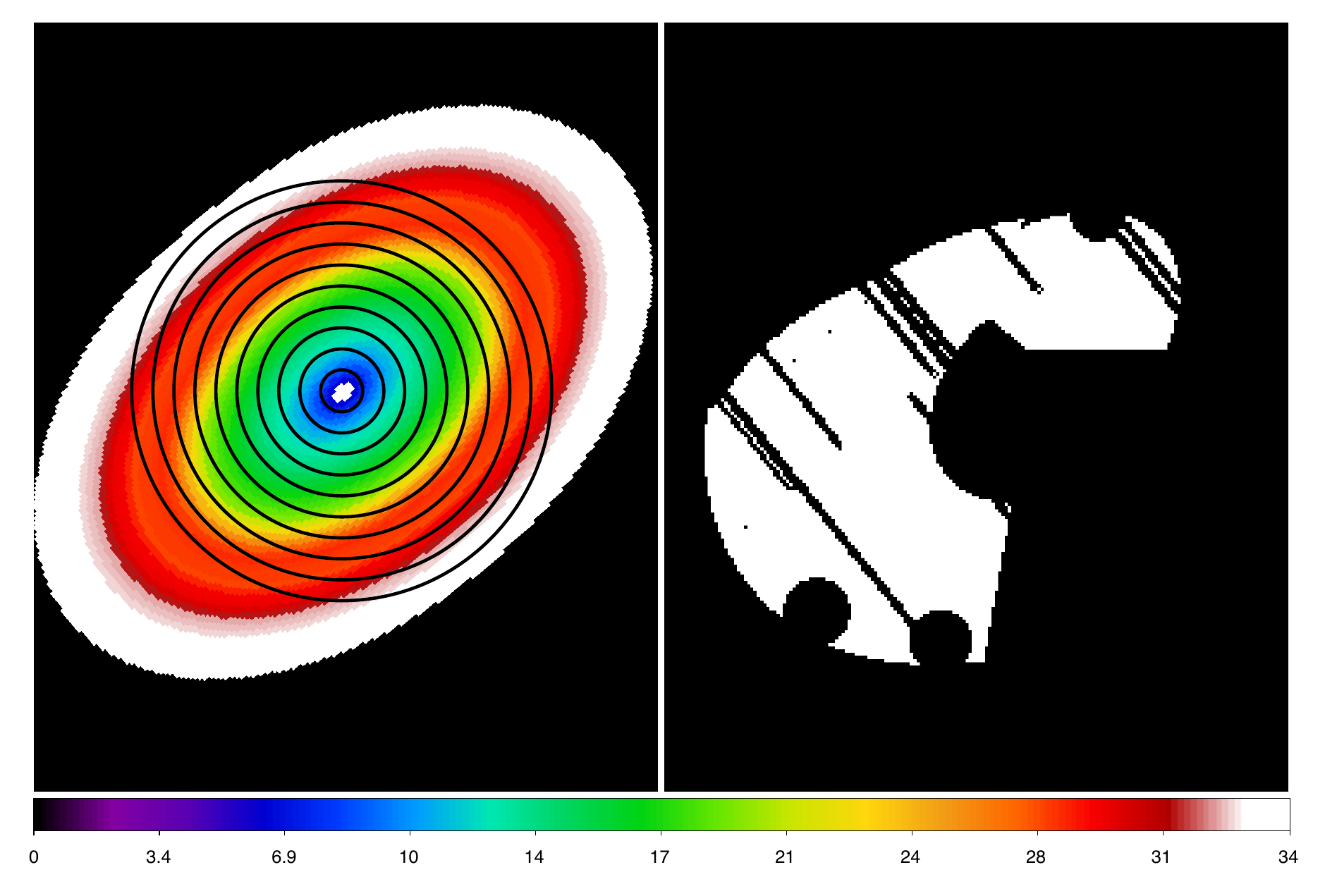}
\caption{Simulated image of an elliptical entropy distribution to which we overlaid circular annuli to qualitatively show the effect of the assumption of spherical symmetry, compared with the mask used to select ``excess'' regions. }
\label{fig:sim}
\end{figure}

We defined a ``sloshing'' region in the residual map (Fig.~\ref{fig:residuals}, right panel) by selecting regions corresponding to brightness excesses). 
We extracted surface brightness profiles in circular annuli, which were intersected with this region, and a temperature profile in spherical sectors, which resemble closely the mask file. We then combined these quantities to make pseudo-entropy profiles, which we show in the left panel of Fig. \ref{entropy}, and compared them with the entropy profile in the ``undisturbed'' complementary region, $K(r)$. 
 As expected, the pseudo-entropy in the excess region is always lower than the mean profile of the cluster. For each radius $r_i$ of the entropy profile, we defined a corresponding radius $c(r_i)\leq r_i$ such that
\begin{equation} K_{excess}(r_i) = K_{unperturbed}(c(r_i)). \end{equation}
Assuming that the gas in the excess region was transported outwards by the sloshing mechanism without modifying its entropy, the radius $c(r)$ thus gives the radius where the gas at radius $r$  was originally located before the sloshing mechanism set in. In Fig.~\ref{entropy} (right panel) we plot the starting radius $c(r)$ as a function of the actual position $r$: the gas is never displaced radially by more than a factor of $\sim 2$, as predicted  by simulations  \citep{ascas06}.\\
We consider a bubble of gas of density $\rho_{in}$ traveling outwards in the potential well of the cluster. At each radius $r$, some gas with density $\rho(r)$ travels inward to fill the bubble volume $V$. The force applied on the bubble is thus given by
\begin{equation}
\vec{F} = V\rho_{in}\vec{g} - V\rho(r)\vec{g} = V\rho_{in}\vec{g}\left(1-\frac{\rho(r)}{\rho_{in}}\right),
\label{eq:force}
\end{equation}
where $\vec{g}$ is the gravity acceleration.
If the process is adiabatic, the bubble expands  at each radius $r$, and the density decreases, preserving the entropy, such that
\begin{equation}K_{in}(r)=P(r)\rho_{in}(r)^{-5/3}=P(r_0)\rho(r_0)^{-5/3}, \end{equation}
where $r_0$ is the initial position of the bubble. We can write its density 
\begin{equation}\rho_{in}(r)=\rho_0\left[\frac{P(r)}{P(r_0)}\right]^{3/5}.\end{equation}
Inserting this into Eq.~\ref{eq:force} and introducing $M_{gas}=\rho_0V$, we find
\begin{eqnarray*}
\vec{F}(r) = M_{gas}\vec{g}(r)\left(1-\frac{\rho(r)}{\rho_{0}}\left[\frac{P(r_0)}{P(r)}\right]^{3/5}\right)=\\ =M_{gas}\vec{g}(r)\left(1-\left[\frac{K(r_0)}{K(r)}\right]^{3/5}\right),
\end{eqnarray*}
where $\vec{g}(r)$ is the gravity acceleration with modulus $GM_{tot}(<r)/r^2$. We calculated the total mass profile using the best fit NFW profile by \citet{ettori02}.
 The total energy needed to move the bubble from radius $r_0$ to $r_1$ is thus given by the integral of the force,
\begin{equation}
E=M_{gas}\int_{r_0}^{r_1} g(r) \left(1-\left[\frac{K(r_0)}{K(r)}\right]^{3/5}\right)\, dr.
\label{eq:de}
\end{equation}
The energy of the sloshing gas is then obtained by summing the contribution of each gas bubble  (Eq.~\ref{eq:de}) for all radii in the excess region:
\begin{equation}
E_{tot}=\sum_{r_i\in \mbox{\scriptsize{excess}}} M_{gas}(r_i)\int_{c(r_i)}^{r_i} g(r) \left(1-\left[\frac{K(c(r_i))}{K(r)}\right]^{3/5}\right)\, dr ,
\label{eq:final}
\end{equation}
where $c(r)$ is the starting radius for each gas particle at radius $r$ that we obtained from the entropy profile. The gas mass $M_{gas}(r_i)$ in Eq.~\ref{eq:final} is calculated as the product of the projected gas density in the $i$-th shell of the profile with the volume, which is defined as the intersection of spherical and cylindrical shells multiplied by the fraction of the azimuth ($f$) considered ($4/3\pi(4r_idr_i)^{3/2}*f$).
Using Eq.~\ref{eq:final}, we find $E_{tot}\sim3.5\times10^{61}$ ergs.\\
In the previous calculation, we estimated the energy spent against the gravitational potential under the assumption of spherical symmetry. This calculation may be biased by the large ellipticity of A2142, inducing an overestimate of the displacement of gas particles and therefore of the energy.
If the entropy distribution  is intrinsically elliptical and we calculate profiles in circular annuli, along the major axis of the ellipse, we will invariably find gas with lower entropy  compared to regions at the same distance from the center but along the minor axis (see Fig.~\ref{fig:sim}, left panel).
To quantify this effect, we generated an image of an elliptical entropy distribution (left panel of Fig.~\ref{fig:sim}): starting from the pseudo-entropy in the ``unperturbed'' region, we used the ellipticity and inclination of the X-ray image of A2142 as geometrical parameters. We then extracted profiles in the ``excess'' and ``unperturbed'' regions, using the mask shown in Fig.~\ref{fig:sim} (right panel) and compared them to measure the displacement $c(r)$. 
As shown in Fig.~\ref{entropy} (left panel, dashed lines), the profile extracted along the ``excess'' region is consistent with the ``unperturbed'' one at small radii, but the two profiles differ significantly at $r>3$ arcmin, although their difference is smaller than the one observed with real data (Fig.~\ref{entropy}). Using these entropy profile, we calculated the displacements induced by the ellipticity, and we subtracted them from the $c(r)$ values calculated before (Fig.~\ref{entropy}, right panel). \\
We note that the correction we calculated in this way may result as an overcorrection, since it assumes that all the ellipticity is intrinsic and not associated to the sloshing itself. For this reason the corrected value should be considered as a lower limit, while the uncorrected one, which completely neglects the ellipticity of the gas, should be considered as a upper limit.

\end{document}